\documentclass{interact}
\usepackage{my}
\usepackage{lscape}
\usepackage[perpage]{footmisc}  
\usepackage[numbers,sort&compress]{natbib}

\newrobustcmd{\Xstate}{\ensuremath{X\,^3\Sigma^-}\xspace}
\newrobustcmd{\Astate}{\ensuremath{A\,^3\Pi}\xspace}
\newrobustcmd{\Bstate}{\ensuremath{B\,^3\Sigma^-}\xspace}
\newrobustcmd{\Cstate}{\ensuremath{C\,^3\Pi}\xspace}
\newrobustcmd{\Cpstate}{\ensuremath{C'\,^3\Pi}\xspace}
\newrobustcmd{\dstate}{\ensuremath{d\,^1\Pi}\xspace}
\newrobustcmd{\astate}{\ensuremath{a\,^1\ensuremath{\Delta}}\xspace}
\newrobustcmd{\AX}[1]{\ensuremath{A(#1)\leftarrow X(0)}\xspace}
\newrobustcmd{\BX}[1]{\ensuremath{B(#1)\leftarrow X(0)}\xspace}
\newrobustcmd{\CX}[1]{\ensuremath{C(#1)\leftarrow X(0)}\xspace}
\newrobustcmd{\dX}[1]{\ensuremath{d(#1)\leftarrow X(0)}\xspace}
\newrobustcmd{\muBX}[1]{\ensuremath{\ensuremath{\mu}_{B-X}}\xspace}
\newrobustcmd{\muCX}[1]{\ensuremath{\ensuremath{\mu}_{C-X}}\xspace}
\newrobustcmd{\wn}[1]{\np[cm^{-1}]{#1}}
\newrobustcmd{\wnfwhm}[1]{\np[cm^{-1}\,FWHM]{#1}}
\newrobustcmd{\abinitio}{{\em ab initio}\xspace}
\renewcommand{\thefootnote}{\textit{\alph{footnote}}}

\begin{document}

\title{Ultraviolet photoabsorption in the $\Bstate-\Xstate$ and $\Cstate-\Xstate$ band systems of SO sulphur isotopologues}

\author{
  \name{A.~N.~Heays,\textsuperscript{a,b,c}
    G.~Stark,\textsuperscript{d}
    J.~R.~Lyons,\textsuperscript{a,e}
    N.~de~Oliveira,\textsuperscript{f}
    B.~R.~Lewis\textsuperscript{g}
    and S.~T.~Gibson\textsuperscript{g}
  }
  \affil{
    \textsuperscript{a}School of Earth and Space Exploration, Arizona State University, Tempe, AZ 85281, USA;
    \textsuperscript{b}NASA Astrobiology Institute, NASA Ames Research Center, Moffett Field, California, USA;
    \textsuperscript{c}J. Heyrovsk\'y Institute of Physical Chemistry, Czech Academy of Sciences, Dolej\v{s}kova 3, CZ18223 Prague 8, Czech Republic
    \textsuperscript{d}Department of Physics, Wellesley College, Wellesley, MA 02481, USA;
    \textsuperscript{e}Planetary Science Institute, Tucson AZ 85719, USA;
    \textsuperscript{f}Synchrotron SOLEIL, L'Orme des Merisiers, D\'epartementale 128, 91190 Saint-Aubin, France;
    \textsuperscript{g}Research School of Physics, The Australian National University, Canberra, ACT 2601, Australia
  }}

\maketitle

\begin{abstract}
  High-resolution far-ultraviolet broadband Fourier-transform photoabsorption spectra of \ce{{}^{32}S {}^{16}O}, \ce{{}^{33}S {}^{16}O}, \ce{{}^{34}S {}^{16}O}, and \ce{{}^{36}S {}^{16}O} are recorded in a microwave discharge seeded with \ce{SO2}.
  The $\Bstate(v=4-30)\leftarrow\Xstate(v=0)$ and $\Cstate(v=0-7)\leftarrow\Xstate(v=0)$ bands are observed or inferred in the \np{43000} to \wn{51000} (196 to 233\,nm) spectral range.
  This is the first experimental detection of a $\Cstate(v>2)$ level and of any of these observed bands in an S-substituted isotopologue.
  Additional measurements of $\Astate(v=1-3)\leftarrow\Xstate(v=0)$ provide a calibration of the SO column density.
  Measured band profiles are fitted to an effective-Hamiltonian model of coupled excited \Bstate and \Cstate states along with their predissociation linewidths and absorption band strengths.
  Electronic-state potential-energy curves and transition moments are deduced.
  The end result is a list of line frequencies, $f$-values, and dissociation widths describing the far-ultraviolet photodissociation spectrum of SO that is accurate enough for computing atmospheric photolytic isotope-fractionation.
\end{abstract}


\section{Introduction}

Sulphur monoxide (SO) occurs with observable abundance in the low-density interstellar medium \citep{gottlieb1973,mateen2006,guilloteau2013,riviere-marichalar2019}, where its spectroscopic signature has been used to investigate the properties of shocked gas and to constrain the ages of star-forming molecular-cloud cores.
It is also a significant species within the Solar system, forming in Io's atmosphere \citep{lellouch1996,moullet2013,de_kleer2019} primarily through the photodissociation of \ce{SO2}, and with detections in cometary comae \citep{bockelee-morvan2000,boissier2007} and the Venus atmosphere \citep{na1990} where a vertical profile was obtained by the Venus Express mission \citep{belyaev2012}. 
SO  and its S-substituted isotopologues may be key molecules in the sulphur cycle of a pre-oxygenated early-Earth atmosphere (older than 2.4\,Gyr) \citep{pavlov2002,ono2017}. 
Measurements of mineral S-isotopes from sedimentary rock reveal fractionating signatures that are likely due to gas-phase photochemistry \citep{farquhar2000a}. 
These signatures are the products of a low-\ce{O2} atmosphere in which sulphur allotropes and elemental sulphur coexist with \ce{H2SO4} and provide a quantitative tracer for the rise of \ce{O2} in Earth's atmosphere.
Additionally, SO and its sulphur isotopologues derived from \ce{SO2} photolysis play a central role in the conversion of oxidised to neutral sulphur \citep{lyons2009,danielache2014}. 
The successful modelling of astrochemical and planetary SO observations, and S-isotopes in the early-Earth atmosphere, requires a full, detailed, and accurate knowledge of the excited states of SO, their transition properties, and temperature- and isotopologue-dependent ultraviolet photoabsorption and photodissociation cross sections. 

\begin{figure}
  \centering
  \includegraphics{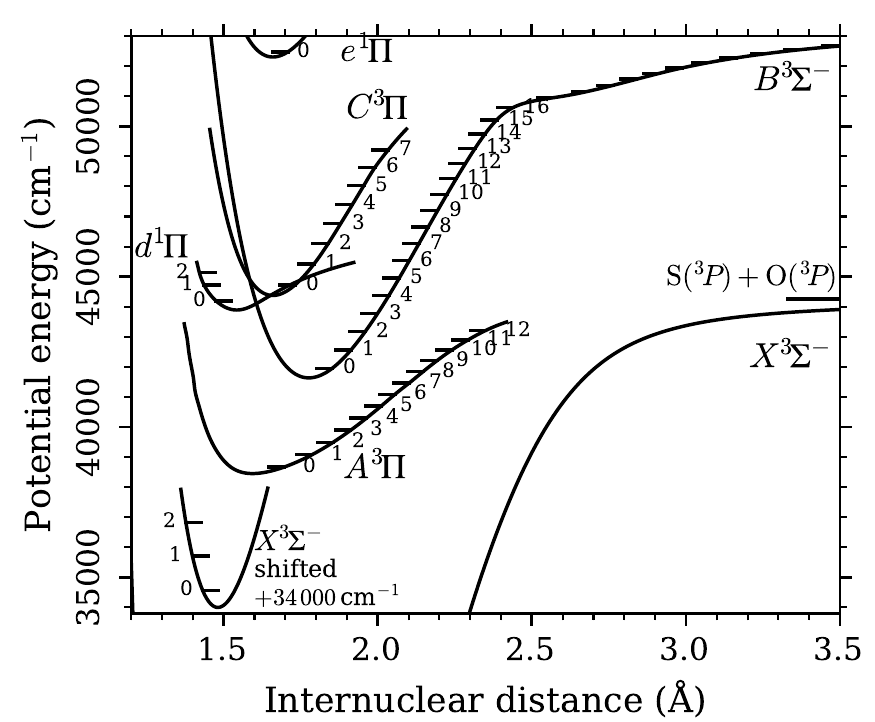}
  \caption{Potential-energy curves of excited and ground states of SO referenced to the ground state equilibrium energy, along with \ce{{}^{32}S {}^{16}O} vibrational level energies.}
  \label{fig:introduction PECs}
\end{figure}

An overview of the experimentally-known photoabsorbing excited states of SO is plotted in Fig.~\ref{fig:introduction PECs}.
The resulting ultraviolet absorption spectrum of SO is dominated by the strong progression of $\Bstate\leftarrow\Xstate$ vibrational bands between 240 and 190\,nm (\np{42000} and \wn{53000}), and is an analogue of the $B\,{}^3\ensuremath{\Sigma}^-_u - X\,{}^3\ensuremath{\Sigma}^-_g$ systems in isovalent \ce{O2} and \ce{S2}.
The early SO spectroscopic study of \citet{martin1932} observed a breaking off in emission from high rotational levels of $\Bstate(v=0-3)$ that is attributable to predissociation.
\citet{clerbaux1994} established the responsible dissociation threshold in a study of emission from low-lying $B(v)$ levels of \ce{{}^{32}S {}^{16}O} and \ce{{}^{32}S {}^{18}O}, and identified multiple local perturbations affecting them.
\citet{colin1969} photographically recorded absorption bands $\Bstate(v')\leftarrow \Xstate(v''=0)$ as far as $v'=30$ using flash-photolysis methods, and provided partial rotational analyses for some bands while noting irregular vibrational spacings and significant line broadening throughout their progression.
A comprehensive study of $\Bstate(v=0-16)$ levels by \citet{liu_ching-ping2006} combined degenerate four-wave mixing, laser-induced fluorescence, Fourier transform emission, and photographic absorption measurements in a detailed rotational analysis of multiple bands appearing between \np{41000} and \wn{50000} and found evidence for numerous perturbations.
They attributed the latter to spin-orbit and rotational interactions between rovibronic levels of the \Bstate, \Astate, \Cstate, and \dstate states by making use of earlier multi-photon ionisation measurements by \citet{elks1999} and \citet{archer2000} characterising the $\Astate(v=0-13)$ and $\Cstate(v=0-1)$ and $\dstate(v=0-3)$ levels, respectively.
\citet{liu_ching-ping2006} estimated upper-limits varying between 0.3 and \wn{40} for the linewidths of $\Bstate(v=4-16)$.

\citet{yamasaki2003b,yamasaki2005} measured radiative lifetimes for $\Bstate(v=0-3)$ and \citet{phillips1981} recorded a low-resolution absorption cross section covering the $\Bstate\leftarrow\Xstate$ progression in a flowing discharge.
Several studies have experimentally measured the radiative lifetime of various \Astate levels and their emission branching between ground state vibrational levels \citep{clyne1982,clyne1986,lo1988,stuart1994,elks1999}.

Early theoretical studies \citep{dixon1977,swope1979b} determined potential-energy curves for the \Xstate, \Bstate, and other low-lying excited states correlated to the two lowest SO dissociation channels, generating $\ensuremath{\ce{S}({}^3{\rm P})}+\ensuremath{\ce{O}({}^3{\rm P})}$ and $\ensuremath{\ce{S}({}^1{\rm D})}+\ensuremath{\ce{O}({}^3{\rm P})}$.
Some of these excited states can, in principle, interact with \Bstate.
In particular, computed \Cstate and \dstate states \citep{ornellas1998b,borin1999,archer2000} adiabatically correlate to $\ensuremath{\ce{S}({}^3{\rm P})}+\ensuremath{\ce{O}({}^3{\rm P})}$, have minima supporting bound vibrational levels, and possess inner and outer limb crossings with \Bstate.
The experimentally-observed variable line broadening of \Bstate suggests that, in addition to any interactions with bound levels of \Astate, \Cstate and \dstate, it  interacts with multiple dissociating states.
Multiple candidate predissociation pathways were noted by \citet{yu2011} following computation of spin-orbit couplings to numerous singlet, triplet, and quintet states.
More recently, potential-energy curves, transition dipole moments, and couplings affecting  \Bstate, and \Cstate have been computed by \citet{danielache2014} and \citet{sarka2019b} for the purposes of studying the isotope-dependence of SO photoabsorption and photodissociation.
\citet{sarka2019b} also compute a higher-lying unbound \ensuremath{{}^3\Sigma^-} state that approaches \Bstate at large internuclear distance and their associated nonadiabatic coupling.
This interaction results in the peculiar shape of the \Bstate potential-energy curve and a sharp alteration of its electronic-configuration near 2.25\,\AA{}, as evidenced by an exchange of electronic transition moment between the diabatic $\Bstate - \Xstate$ and higher-energy $\ensuremath{{}^3\Sigma^-} - \Xstate$ transitions.
Further \abinitio calculations of excited SO states have been recently performed by \citet{feng2019b}, who also computed many spin-orbit interaction energies, and \citet{da_silva2020}.
Previous experimental studies \citep{colin1969,clerbaux1994} concerning the photodissociation spectrum of SO isotopologues appear to be limited to \ce{{}^{32}S {}^{18}O} with theoretical results extended to all S substitutions \citep{sarka2019b}.

In this report, we present spectroscopic analyses of high-resolution broadband absorption spectra of the $\Bstate(v'=4-30)\leftarrow\Xstate(v''=0)$ and $\Cstate(v'=0-7)\leftarrow\Xstate(v''=0)$ systems for four SO isotopologues: \ce{{}^{32}S {}^{16}O}, \ce{{}^{33}S {}^{16}O}, \ce{{}^{34}S {}^{16}O}, and \ce{{}^{36}S {}^{16}O}.
The observed absorption band profiles are reduced to deperturbed molecular constants and spin-orbit interaction  energies mixing \Bstate and \Cstate electronic-vibrational states, along with calibrated transition moments with the ground state and a quantification of the observed predissociation line broadening.
Empirical \Bstate and \Cstate potential-energy curves and a global spin-orbit interaction are fitted to these data and used to extrapolate the experimental data to include all bands up to the \Bstate dissociation limit in all isotopologues.

\section{Measurements}

Photoabsorption measurements were performed on the high-resolution absorption spectroscopy branch of the DESIRS beamline \citep{nahon2012} at the SOLEIL synchrotron.
This facility was used in similar studies of the OH and \ce{S2} radicals \citep{heays2018a,stark2018}.
The beamline undulator generates continuum bandpass radiation with a width of 7\% of its central frequency.
Four overlapping measurements were required for complete coverage of the \np{43000} to \wn{52000} target region.
The continuum radiation was passed through a rare-gas-filled chamber to filter unwanted higher harmonics generated in the undulator, then an absorption cell, and terminated at a vacuum-ultraviolet Fourier-transform spectrometer \citep{de_oliveira2011,de_oliveira2016}.
This is an all-reflection wave-front-division interferometer reliant on spatial coherence of the synchrotron beam and a modified Fresnel bi-mirror configuration with the optical-path difference scanned by translating one reflector.
The instrument was operated with spectral resolution between 0.15 and \np[cm^{-1}]{0.86} full-width at half-maximum (FWHM) depending on the perceived sharpness of SO features in each undulator bandpass and signal-to-noise considerations.

SO radicals were produced in a flowing discharge containing one of four sulphur dioxide samples seeded in helium.
Highly enriched \ce{^{33}SO2} (99\% \ce{{}^{33}S}), \ce{^{34}SO2} (99.8\% \ce{{}^{34}S}), and \ce{^{36}SO2} (approximately 70\% \ce{{}^{36}S}, 20\% \ce{{}^{34}S}, and 10\% \ce{{}^{32}S}) gases were used to generate rare SO isotopologues, and natural abundance \ce{SO2} (95.02\% \ce{{}^{32}S}, 0.75\% \ce{{}^{33}S}, and 4.21\% \ce{{}^{34}S}) was used to study \ce{{}^{32}S {}^{16}O}.
All three samples contained oxygen in natural abundance but no absorption due to \ce{{}^{18}O}-bearing species was detected.
A radio-frequency generator (13.5\,MHz, 200\,W power) was centred in a 1.5\,m glass absorption cell equipped with wedged \ce{MgF2} windows.
Helium carrier gas with an upstream pressure of 2\,mbar flowed through the cell and was continuously evacuated by a \np[m^3\,hr^{-1}]{600} Roots pump.
Fluorescence in the discharge typically extended 40\,cm on either side of the central generator cavity.
\ce{SO2} was seeded into the He flow prior to the absorption cell with a partial pressure between 0.04 and 0.1\,mbar.
The cell flow rate was significantly reduced when recording \ce{{}^{33}S {}^{16}O}, \ce{{}^{34}S {}^{16}O}, and \ce{{}^{36}S {}^{16}O} spectra to minimise the consumption of rare \ce{SO2} isotopologues.

Strong absorption features of the parent \ce{SO2} $\tilde C\,{}^1B_2-\tilde X\,{}^1A_1$ electronic system \citep{freeman1984,stark1999b} between \np{45500} and \wn{57000} overlap most SO \BX{v} absorption bands.
A sequence of three spectra were recorded for each bandpass.
First, the synchrotron radiation bandpass was established in a spectrum of pure helium.
\ce{SO2} was then seeded into the He flow and a reference absorption spectrum recorded with the discharge off.
Finally, a combined SO and \ce{SO2} spectrum was recorded after activating the discharge source.
The raw experimental spectra are available in an online archive \citep{online_appendix}.

\section{Analysis method}
\label{sec:analysis method}

The \Xstate and \Bstate states of SO consist of three Hund's case-(a) spin-, $e/f$-parity sublevels, identifiable, in order of increasing energy, with the following quantum numbers:
\begin{align*}
  (F_1,e) &:\ensuremath{\Lambda}=0,\ \ensuremath{\Omega}=0,\ \ensuremath{\Sigma}=0\,; \\
  (F_2,f) &:\ensuremath{\Lambda}=0,\ \ensuremath{\Omega}=1,\ \ensuremath{\Sigma}=1\,; \\  
  \text{and }(F_3,e) &:\ensuremath{\Lambda}=0,\ \ensuremath{\Omega}=1,\ \ensuremath{\Sigma}=1,
\end{align*}
while \Astate consists of six sublevels: 
\begin{align*}
  (F_1,e/f) &:\ensuremath{\Lambda}=1,\ \ensuremath{\Omega}=0,\ \ensuremath{\Sigma}=-1\,; \\
  (F_2,e/f) &:\ensuremath{\Lambda}=1,\ \ensuremath{\Omega}=1,\ \ensuremath{\Sigma}=0\,; \\
  \text{and }(F_3,e/f) &:\ensuremath{\Lambda}=1,\ \ensuremath{\Omega}=2,\ \ensuremath{\Sigma}=+1,
\end{align*}
and higher-$\ensuremath{\Omega}$ levels of \Cstate occur with lower energy:
\begin{align*}
  (F_1,e/f) &:\ensuremath{\Lambda}=1,\ \ensuremath{\Omega}=2,\ \ensuremath{\Sigma}=+1\,; \\
  (F_2,e/f) &:\ensuremath{\Lambda}=1,\ \ensuremath{\Omega}=1,\ \ensuremath{\Sigma}=0\,; \\
  \text{and }(F_3,e/f) &:\ensuremath{\Lambda}=1,\ \ensuremath{\Omega}=0,\ \ensuremath{\Sigma}=-1.
\end{align*}
Here, $\ensuremath{\Lambda}$ and $\ensuremath{\Sigma}$ are the usual electronic-orbital and spin angular momentum projection quantum numbers, and $\ensuremath{\Omega}=|\ensuremath{\Lambda}+\ensuremath{\Sigma}|$.
States of common parity within each $F_i$ manifold become mixed with increasing molecular rotation, although relatively slowly in the cases of \Astate and \Cstate because of their large spin-orbit splittings.
In principle, $\Bstate(v')\leftarrow\Xstate(v'')$ bands consist of 14 overlapping rotational branches but with reduced contributions from spin-forbidden $\ensuremath{\Delta}\ensuremath{\Sigma}\neq 0$ transitions.
The rotational structure of $\Astate(v')\leftarrow\Xstate(v'')$ and $\Cstate(v')\leftarrow\Xstate(v'')$ bands consists of 27 branches, but transitions terminating on different \Cstate $\ensuremath{\Omega}$-substates are well separated in energy.
In what follows, fully-specified electronic-vibrational states and transitions are sometimes abbreviated, e.g., $\Xstate(v=0)$ to $X(0)$, and $\Cstate_{\ensuremath{\Omega}=0}(v'=1)\leftarrow\Xstate(v''=0)$ to $C(v=1,\ensuremath{\Omega}=0)\leftarrow X(0)$.

\subsection{Vibrational state model}
\label{sec:vibrational state model}

The significant predissociation broadening of nearly all \BX{v'} bands and poor signal-to-noise of the observed \CX{v'} absorption precludes their line-by-line analysis in most cases.
Instead, the ground and excited vibrational levels are modelled with a minimal set of molecular parameters in a Hund's case-(a) parity-symmetrised basis, and consideration is made for spin-rotational mixing of $\ensuremath{\Omega}$-sublevels within each electronic-vibrational state and for spin-orbit mixing of neighbouring $B(v_B)$ and $C(v_C)$ levels.

An effective-Hamiltonian matrix is built with diagonal deperturbed rotational level energies and off-diagonal interaction energies.
Diagonalising this matrix produces a model of observable energies and mixing coefficients for the interacting case-(a) levels.
The form, symbols, and phase conventions adopted in our matrix diagonalisation and line strength calculation are the same as those used for linear molecules by the PGOPHER program \citep{western2017}, that is, the effective Hamiltonian of \citet{brown1979} with explicit matrix elements for \ensuremath{{}^3\Sigma^-} and \ensuremath{{}^3\Pi} spin-rotation-mixed manifolds listed in \citet{cheung1986} and \citet{brown_merer1979}, respectively.
The $o$ and $p$ $\ensuremath{\Lambda}$-doubling terms of \citet{brown_merer1979} were used in the analysis of some \Cstate and \Astate levels.

The interaction of $\Bstate(v_B)$ and $\Cstate(v_C)$ levels is modelled as a spin-orbit mixing of levels differing by $\ensuremath{\Delta}\ensuremath{\Sigma}=\pm1$, $\ensuremath{\Delta}\ensuremath{\Lambda}=\mp1$, and $\ensuremath{\Delta}\ensuremath{\Omega}=0$, and with common $e/f$ symmetry.
A reduced matrix element \citep{lefebvre-brion_field2004} is fitted to each pair of interacting $B(v_B)$ and $C(v_C)$ levels following the definition and phase convention of PGOPHER. 
This reduced matrix element is $-\sqrt{6}$ times larger than the conventionally-referenced matrix element between $\ensuremath{\Omega}=1$ levels:
\begin{equation}
  \label{eq:specific spin orbit matrix elements Omega1}
  \mybraopket{\Bstate_{\ensuremath{\Omega}=1,e/f}}{H^\text{SO}}{\Cstate_{\ensuremath{\Omega}=1,e/f}} = \xi_{v_Bv_C},
\end{equation}
that we quote below.
The corresponding matrix element mixing $\ensuremath{\Omega}=0$ levels is
\begin{equation}
  \label{eq:specific spin orbit matrix elements Omega0}
  \mybraopket{\Bstate_{\ensuremath{\Omega}=0,e/f}}{H^\text{SO}}{\Cstate_{\ensuremath{\Omega}=0,e/f}} = \sqrt{2}\xi_{v_Bv_C}.
\end{equation}

Fitted scalar electric-dipole vibronic transition moments, $\ensuremath{\mu}_{B(v)-X(0)}$ and $\ensuremath{\mu}_{C(v)-X(0)}$, are used to simulate the observed $\ensuremath{{}^3\Sigma^-}\leftarrow\ensuremath{{}^3\Sigma^-}$ and $\ensuremath{{}^3\Pi}\leftarrow\ensuremath{{}^3\Sigma^-}$ absorption bands.
The transition moments for parity-allowed $\ensuremath{\Delta}J=-1,0$ and $+1$ rovibronic transitions between unmixed case-(a) levels
are computed according to \citep{hougen1970,lefebvre-brion_field2004,hansson2005}:
\begin{multline}
  \label{eq:rotational transition moments}
  \ensuremath{\mu}_{i(v'J'\ensuremath{\Omega}')-X(v''J''\ensuremath{\Omega}'')} = \ensuremath{\mu}_{i(v')-X(v'')}\sqrt{(2J''+1)(2J'+1)}\\
  \times  (-1)^{J'-\ensuremath{\Omega}'+\ensuremath{\delta}_{\ensuremath{\Lambda}'0}}
  \begin{pmatrix} J' & 1 &J'' \\ -\ensuremath{\Omega}' & \ensuremath{\Omega}'-\ensuremath{\Omega}'' &\ensuremath{\Omega}'' \end{pmatrix},
\end{multline}
where the final term is a Wigner-$3j$ coefficient.
The mixing coefficients calculated while diagonalising the excited and ground state level energies are used to compute mixed line strengths which should match the observed transition strengths.
The mixed absorption $f$-values are
\begin{equation}
  f = \np{3.038e-6}\times \ensuremath{\nu}_0 \Bigl(\ensuremath{\mu}^\text{mixed}_{i(v'J'\ensuremath{\Omega}')-X(v''J''\ensuremath{\Omega}'')}\Bigr)^2,
\end{equation}
where the given numerical constant assumes $\ensuremath{\mu}$ in atomic units and a transition wavenumber, $\ensuremath{\nu}_0$, with units of cm$^{-1}$.

Lorentzian predissociation line broadening is modelled by adopting complex-valued diagonal level energies in the effective Hamiltonian, where the imaginary component is equivalent to the FWHM linewidth, $\ensuremath{\Gamma}$.
Aside from conveniently accounting for mixed linewidths, the adoption of complex level energies also slightly modifies the computed lineshifts due to level interactions between nearby case-(a) levels with overlapping line wings.
The observed line broadening is caused by repulsive states not included in our deperturbation matrix, and the broadening of each deperturbed electronic-vibrational levels is experimentally determined.
In some cases, $\ensuremath{\Omega}$- and $J$-dependent widths of case-(a) states are required to reproduce the experimental spectra.

\begin{figure*}
  \centering
  \includegraphics[width=\columnwidth]{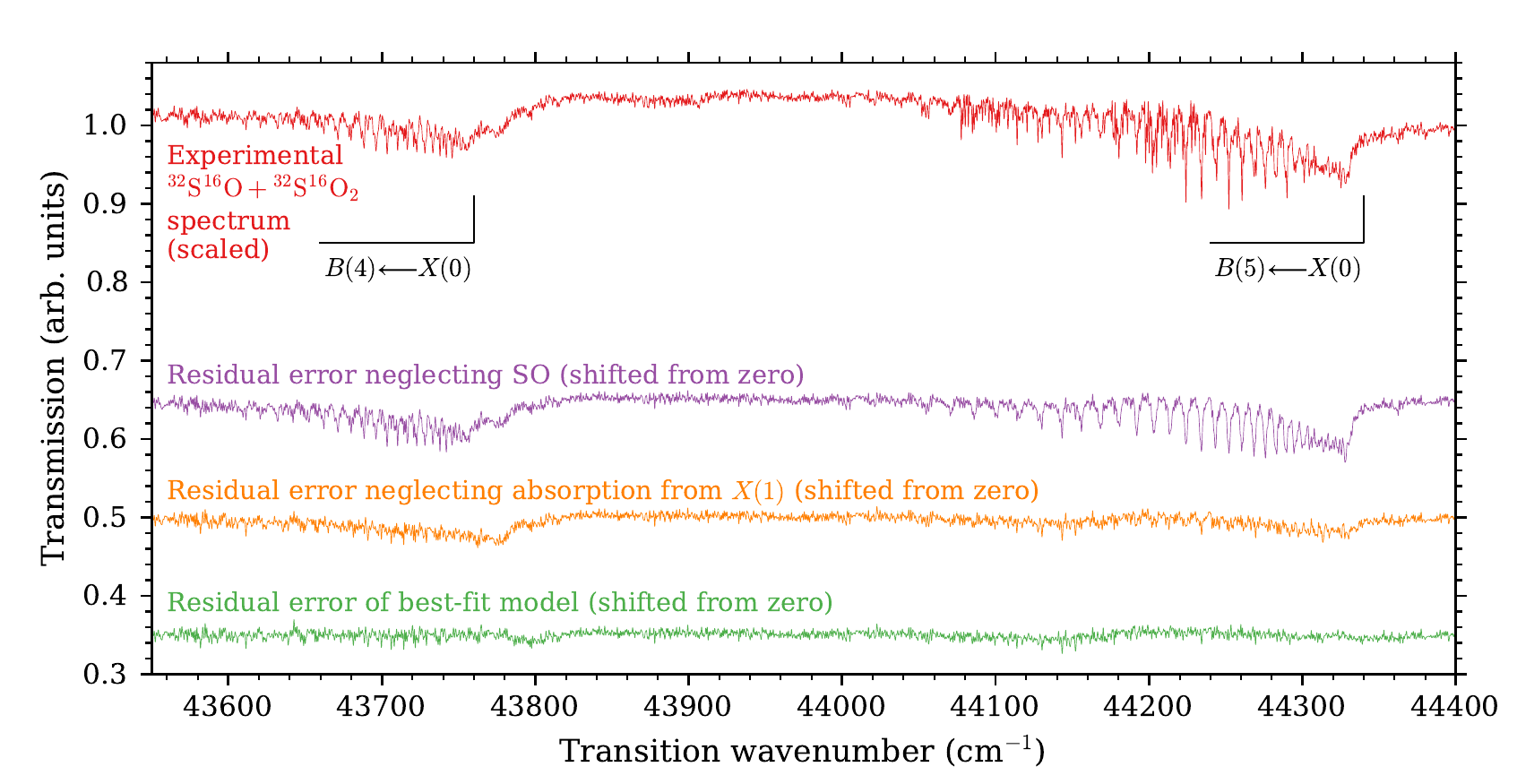}
  \caption{An experimental spectrum showing $\ce{{}^{32}S {}^{16}O}\,\BX{4}$ and \BX{5} absorption and the residual error of a best-fit model of these bands. Residual errors are also shown for models neglecting all \ce{{}^{32}S {}^{16}O} absorption and absorption from vibrationally-excited $X(v=1)$ levels.}
  \label{fig:spectrum B04-X00}
  \label{fig:spectrum B05-X00}
\end{figure*}

\begin{figure*}
  \centering
  \includegraphics[width=\columnwidth]{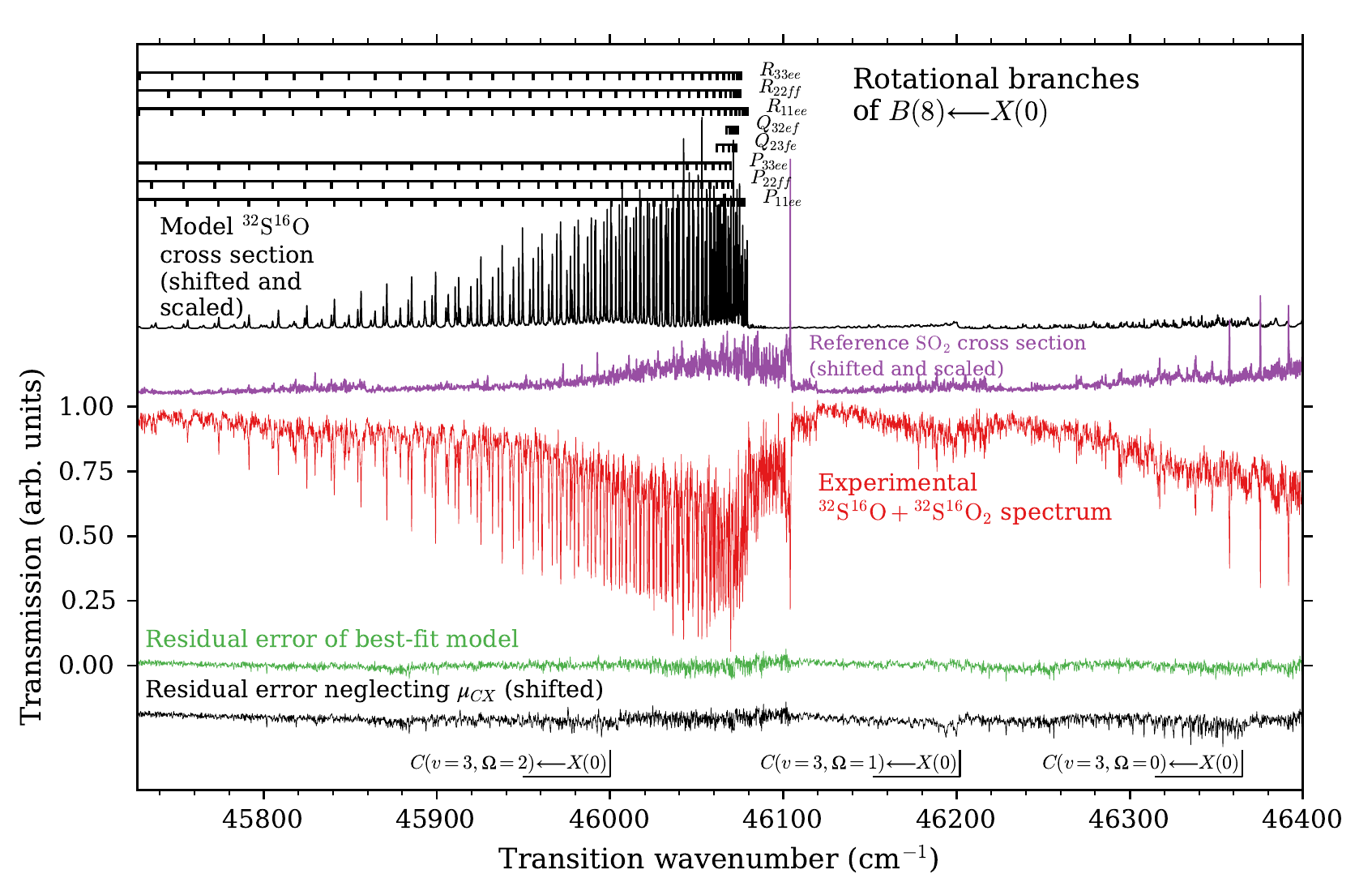}
  \caption{A rotationally-assigned experimental spectrum showing \ce{{}^{32}S {}^{16}O} \BX{8} and \CX{3} absorption. The plotted reference \ce{SO2} and modelled SO cross sections were used to synthesise a spectrum with residual error as shown. Also shown is the residual error of a model neglecting the $\ensuremath{\mu}_{C(v=3)-X(v=0)}$ transition moment.}
  \label{fig:spectrum B08-X00}
\end{figure*}

\begin{figure*}
  \centering
  \includegraphics[width=\columnwidth]{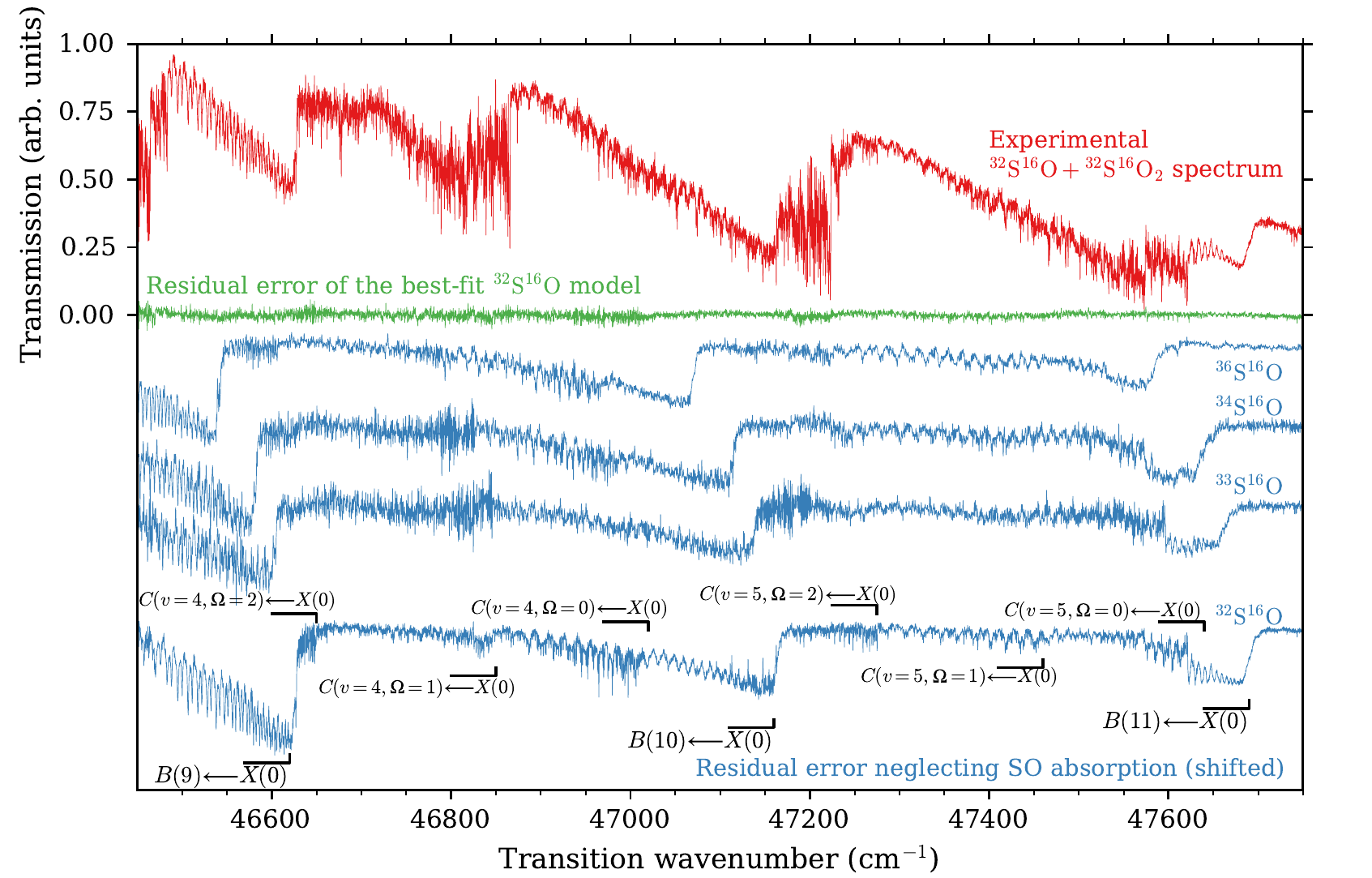}
  \caption{An experimental spectrum showing \ce{{}^{32}S {}^{16}O} \BX{9}, \BX{10}, \BX{11}, \CX{4}, and \CX{5} absorption and the residual error of a best-fit model labelled with bandhead assignments. Also shown are residual errors of similar models for other S-substituted isotopologues.}
  \label{fig:spectrum B09-X00}
  \label{fig:spectrum B10-X00}
  \label{fig:spectrum B11-X00}
  \label{fig:spectrum C04-X00}
  \label{fig:spectrum C05-X00}
\end{figure*}

A list of perturbed line frequencies, $\ensuremath{\nu}_i$, $f$-values, $f_i$, and linewidths, $\ensuremath{\Gamma}_i$, was computed for rotational lines with $J''\leq 50$ for all observed bands.
A cross section composed of all transitions was computed assuming a thermal population distribution of ground-state rotational levels, $\ensuremath{\alpha}_{F''_iJ''}$ and a Voigt line profile, $V(\ensuremath{\nu},\ensuremath{\Gamma},\ensuremath{\Gamma}_\text{D})$, constructed assuming a Gaussian Doppler width, $\ensuremath{\Gamma}_\text{D}$,
\begin{equation}
  \ensuremath{\sigma}(\ensuremath{\nu}) = \sum_i \np{8.853e-13} × f_i\ensuremath{\alpha}_{F''_iJ''_i}V(\ensuremath{\nu}-\ensuremath{\nu}_{0i},\ensuremath{\Gamma}_i,\ensuremath{\Gamma}_\text{D}),
\end{equation}
where the numerical constant assumes a cross section in units of cm$^2$.

The synchrotron radiation generated by the beamline undulator possesses a peaked band pass of approximately 5\,nm FWHM and its curvature, $I_0(\ensuremath{\nu})$, was modelled with a spline function with knots separated by \wn{500}.
The Beer-Lambert law was used to compute an ideal absorption spectrum:
\begin{equation}
  I(\ensuremath{\nu}) = I_0(\ensuremath{\nu})\exp{\Bigl(-N_{\ce{SO}}\ensuremath{\sigma}(\ensuremath{\nu}) - \tau_\text{other}(\ensuremath{\nu})\Bigr)},
\end{equation}
assuming a column density of SO radicals, $N_{\ce{SO}}$, and including extra opacity due to contaminant absorbers in the spectrum, $\tau_\text{other}$.
The ideal spectrum was convolved with a function defining the instrumental resolution, primarily a sinc function with a small amount of additional broadening previously found to be generated by this instrument \citep{heays2011b} and modelled as a Gaussian of \wnfwhm{0.1}.
This final spectrum is compared pointwise with the measured spectra and the various parameters governing the model were adjusted iteratively until a best agreement was found in the least-squares sense.
Examples of this comparison for several \BX{v} bands are plotted in Figs.~\ref{fig:spectrum B04-X00}, \ref{fig:spectrum B08-X00}, and \ref{fig:spectrum B09-X00} and include spectra exhibiting varied \ce{SO2} contamination and line confusion due to predissociation broadening.
Figure~\ref{fig:spectrum B08-X00} is annotated to show the full rotational structure of \BX{8}, including all rotational transitions stronger than 2\% of the most absorbed.
The same structure underlies the unresolved $B-X$ bands.

The overlapping absorption of parent-gas \ce{SO2} isotopologues was comparable to that of SO itself and high-spectral-resolution \ce{SO2} cross sections measured for the purpose were included in our modelling.
An increase of \ce{SO2} rotational temperature apparently occurs when the discharge is struck so the reference spectra do not quite account for all \ce{SO2} absorption near its bandheads.
The recorded spectra of \ce{{}^{36}S {}^{16}O} were contaminated by \ce{{}^{32}S {}^{16}O} and \ce{{}^{34}S {}^{16}O} beyond the expected amount given the known impurity of the \ce{^{36}SO2} parent gas.
These spectra were recorded subsequent to and on the same apparatus as other studies of \ce{^{32}S}-  and \ce{^{34}S}-containing molecules, and outgassing of earlier-deposited sulphur might explain the extra contamination.
Ultimately the mixture of SO isotopologues was assessed directly from the spectra and has a ratio near $\ce{^{36}S}{\,:\,}\ce{^{34}S}{\,:\,}\ce{^{32}S}=1{\,:\,}0.28{\,:\,}0.16$, with some variation between measurements.
Additional SO absorption from super-thermally excited $\Xstate(v=1)$ and electronically-excited $\astate(v=0)$ was also recorded and accounted for with additional vibrational state models.
Significant absorption due to $B(v'+2)\leftarrow{}X(1)$ hot bands was found to overlap $B(v')\leftarrow{}X(0)$ for $v'=5-8$ in all isotopologues.

Multiple overlapping spectra were recorded to span the desired spectral range and some additional spectra were recorded under varied discharge conditions.
Up to five independent spectra were recorded for some bands and the parameters governing their upper-state levels and transition moments were fitted simultaneously.
This effectively increased the signal-to-noise-ratio for these bands.

Many trials were necessary before arriving at a set of well-fitting and physically-reasonable deperturbed level energies and $\Bstate\sim\Cstate$ interaction parameters, due to the non-resolution of most \BX{v} rotational structure and the weakness or non-observation of \CX{v} bands.
The inclusion of higher-order centrifugal-distortion parameters (for example $H$, $\ensuremath{\lambda}_D$, $\ensuremath{\gamma}_D$, and $A_D$) was found mostly unnecessary during this process and some small but significant parameters (for example $D$ and $\ensuremath{\gamma}$) were set to fixed values for some bands once their overall pattern-forming values had been determined.
Statistical fitting uncertainties of the model parameters are estimated by the least-squares fitting routine but do not account for correlation between model parameters or reflect model error introduced by the assumption of a particular effective Hamiltonian, constraints imposed on its parameters, or the incomplete deperturbation of any level.
The scatter of our fitted parameters significantly exceeds these model parameter uncertainties.
Therefore, in this work we prefer to list estimated uncertainties more reflective of the experimental scatter.
However, the listed uncertainty estimates are best considered as relative.

An SO rotational temperature of $360\pm{}15$\,K was determined from the strengths of rotational lines in the well-resolved \BX{8} band, and was used to define the model Doppler width, $\Gamma_D$.
A thermalised population of ground-state levels peaks near $J=13$ at 360\,K, falls below 5\% of its peak value by $J=40$, and consists of 98\% $v=0$.

Sufficient overlap exists between spectra to calculate relative column densities separately for each isotopologue.
A calibration of the \ce{{}^{32}S {}^{16}O} column-density, and $B-X$ and $C-X$ transition moments, was determined from a measurement of well-known $A(v)-X(0)$ transitions.
A calibration of transition moments of other isotopologues was made relative to \ce{{}^{32}S {}^{16}O} assuming an isotopologue-independent summation of \BX{v=7-14} transition intensity.

\subsection{Electronic state model}
\label{sec:electronic state model}

The band-by-band analysis of spectroscopic constants described above is necessary for analysing the blended spectra.
Each single-band model, or model of a few interacting bands, results in a computed rotational line list that provides, within experimental uncertainty, the frequencies, strengths, and widths of all unresolved lines that significantly contribute to the experimental band profiles.
A more fundamental model of \Xstate, \Bstate, and \Cstate potential-energy curves;  \Bstate and \Cstate spin-orbit interaction mixing; and electronic transition moments controlling $\Bstate\leftarrow\Xstate$ and $\Cstate\leftarrow\Xstate$ absorption; was fitted to the band-by-band rotational line lists.

\begin{table}
    \caption{Matrix elements of spin-rotation mixed spin-electronic states, $\myket{\ensuremath{\Omega},e/f}$.\textsuperscript{a}}
    \label{tab:electronic state matrix elements}
    \begin{minipage}{\linewidth}
    \centering
    \small
    \begin{tabular}{c@{=\,}l}
      \multicolumn{2}{c}{\Large\ensuremath{{}^3\Sigma^-}:} \\
      \mybraopket{{0,e}}{H}{{0,e}}        & $  T(R)+B(R)\left[J(J+1)+2\right]-2\ensuremath{\gamma}-\frac{4}{3}\ensuremath{\lambda}              $ \\
      \mybraopket{{1,e/f}}{H}{{1,e/f}}& $  T(R)+B(R)J(J+1)-\ensuremath{\gamma}+\frac{2}{3}\ensuremath{\lambda}                  $ \\
      \mybraopket{{0,e}}{H}{{1,e}}        & $  2\sqrt{J(J+1)}\left[-B(R)+\frac{1}{2}\ensuremath{\gamma}\right] $ \\[3ex]
      \multicolumn{2}{c}{\Large\ensuremath{{}^3\Pi}:} \\
      \mybraopket{{0,e/f}}{H}{{0,e/f}}        &$ T(R)+B(R)\left[J(J+1)+2\right]-2\ensuremath{\gamma}-\frac{4}{3}\ensuremath{\lambda}            $ \\
      \mybraopket{{1,e/f}}{H}{{1,e/f}}        &$ T(R)+A+B(R)\left[J(J+1)-2\right]+\frac{2}{3}\ensuremath{\lambda}            $ \\
      \mybraopket{{2,e/f}}{H}{{2,e/f}}        &$ T(R)-A+B(R)\left[J(J+1)+2\right]-2\ensuremath{\gamma}+\frac{2}{3}\ensuremath{\lambda}      $ \\
      \mybraopket{{0,e/f}}{H}{{1,e/f}}        &$ \sqrt{J(J+1)}\left[-\sqrt{2}B(R)+\frac{1}{\sqrt{2}}\ensuremath{\gamma}\right]    $ \\
      \mybraopket{{0,e/f}}{H}{{2,e/f}}        &$ 0                                               $ \\
      \mybraopket{{1,e/f}}{H}{{2,e/f}}        &$ \left[-\sqrt{2}B(R)+\frac{1}{\sqrt{2}}\ensuremath{\gamma}\right]\sqrt{J(J+1)-2}$ \\
    \end{tabular}
    \footnotetext[1]{$B(R)=\frac{\hbar^2}{2\ensuremath{\mu}R^2}$}
  \end{minipage}
\end{table}
Parameterised forms for the diabatic \Bstate, and \Cstate potential-energy curves, $T(R)$ where $R$ is the internuclear-distance, are discussed in Sec.~\ref{sec:potential-energy curves}. 
Separate curves for $\ensuremath{\Omega}$ and $e/f$-parity substates were generated assuming $R$-independent spin-orbit and spin-spin interaction parameters, $A$ and $\ensuremath{\lambda}$, respectively.
To simulate a rotating molecule, $R-$ and reduced-mass-dependent diagonal and off-diagonal matrix elements were computed to represent centrifugal effects and the spin-rotation mixing of $\ensuremath{\Omega}$ levels within each electronic state.
The specific matrix elements used are given in Table~\ref{tab:electronic state matrix elements} and follow the formulation of \citet{brown1979}.

A ladder of uncoupled vibrational energy eigenvalues and wavefunctions, $\ensuremath{\chi}_{i(vJ)}$, was computed from the potential-energy curve of each electronic state for a range of $J$, using the Numerov method \citep{johnson1977}.
Band transition moments between unmixed electronic-vibrational states were computed according to: 
\begin{equation}
  \label{eq:vibrational transition moment}
  \ensuremath{\mu}_{i(v'J')-X(v''J'')} = \ensuremath{\mu}_{i-X}\int_0^\infty \ensuremath{\chi}_{i(v'J')}(R)\ensuremath{\chi}_{X(v''J'')}(R)\,dR,
\end{equation}
where transitions are only allowed between states of common $\ensuremath{\Sigma}$, $i$ represents either the $B$ or $C$ state, and $\ensuremath{\mu}_{i-X}$ is an isotopologue- and $R$-independent electronic transition moment.

Spin-orbit reduced matrix elements mixing all neighbouring and remote $B(v_B)$ and $C(v_C)$ levels were  computed from a fitted scalar parameter, $\ensuremath{\xi}_{BC}$, according to:
\begin{equation}
  \label{eq:vibrational spin orbit energy}
  \ensuremath{\xi}_{v_Bv_CJ} = \ensuremath{\xi}_{BC}\int_0^\infty \ensuremath{\chi}_{B(v_B,J)}(R)\,\ensuremath{\chi}_{C(v_C,J)}(R)\,dR,
\end{equation}
and specific $\ensuremath{\Omega}'\sim \ensuremath{\Omega}''$ matrix elements were computed as in Eqs.~(\ref{eq:specific spin orbit matrix elements Omega1}) and (\ref{eq:specific spin orbit matrix elements Omega0}).

For each value of $J$, full sets of ground- and excited-state uncoupled vibrational energy levels and the spin-orbit interaction energies mixing them were diagonalised.
From the mixed levels a spectrum for all optically-allowed rotational transitions was computed.
The parameters of this global model are mass-independent and simultaneously constrained by all isotopologue measurements.

In the following analysis we also employ band-integrated $f$-values that neglect $B\sim C$ spin-orbit coupling computed according to \citep{larsson1983}:
\begin{equation}
  \label{eq:band fvalue}
  f_{i(v')-X(v'')} = \np{3.038e-6}\times \ensuremath{\nu} \ensuremath{\mu}_{i(v'0)-X(v''0)}^2\frac{2-\ensuremath{\delta}_{0,\ensuremath{\Lambda}'+\ensuremath{\Lambda}''}}{2-\ensuremath{\delta}_{0,\ensuremath{\Lambda}''}},
\end{equation}
and band-averaged emission rates (s$^{-1}$) computed according to
\begin{align}
  \label{eq:band emission rate}
  A_{i(v')-X(v'')} = \np{2.026e-6}\times \ensuremath{\nu}^3 \ensuremath{\mu}_{i(v'0)-X(v''0)}^2\frac{2-\ensuremath{\delta}_{0,\ensuremath{\Lambda}'+\ensuremath{\Lambda}''}}{2-\ensuremath{\delta}_{0,\ensuremath{\Lambda}'}}.
\end{align}

\begin{landscape}
  \setcounter{footnote}{0}
  \setlength\tabcolsep{2pt}
  \renewcommand{\thefootnote}{\textit{\alph{footnote}}}

  \begin{small}
\begin{longtable}{lllrlllllllc}
      \caption{Deperturbed molecular constants and predissociation broadening parameters.\protect\footnotemark}\\
      \label{tab:level parameters}\\
      \toprule
      \multicolumn{1}{c}{Level}          & \multicolumn{1}{c}{$T$\footnotemark} & \multicolumn{1}{c}{$B$} & \multicolumn{1}{c}{$D$}   & \multicolumn{1}{c}{$A$} & \multicolumn{1}{c}{$\ensuremath{\lambda}$} & \multicolumn{1}{c}{$\ensuremath{\gamma}$} & \multicolumn{1}{c}{$\ensuremath{\Gamma}(\text{all-}\ensuremath{\Omega})$\footnotemark} & \multicolumn{1}{c}{$\ensuremath{\Gamma}(\ensuremath{\Omega}=0)$} & \multicolumn{1}{c}{$\ensuremath{\Gamma}(\ensuremath{\Omega}=1)$} & \multicolumn{1}{c}{$\ensuremath{\Gamma}(\ensuremath{\Omega}=2)$} & \multicolumn{1}{c}{Obs. $\ensuremath{\Omega}$\footnotemark} \\
      \midrule
      \endfirsthead
      \multicolumn{11}{l}{Table continued from previous page.}\\
      \toprule
      \multicolumn{1}{c}{Level}          & \multicolumn{1}{c}{$T$} & \multicolumn{1}{c}{$B$} & \multicolumn{1}{c}{$D$}   & \multicolumn{1}{c}{$A$} & \multicolumn{1}{c}{$\ensuremath{\lambda}$} & \multicolumn{1}{c}{$\ensuremath{\gamma}$} & \multicolumn{1}{c}{$\ensuremath{\Gamma}(\text{all-}\ensuremath{\Omega})$} & \multicolumn{1}{c}{$\ensuremath{\Gamma}(\ensuremath{\Omega}=0)$} & \multicolumn{1}{c}{$\ensuremath{\Gamma}(\ensuremath{\Omega}=1)$} & \multicolumn{1}{c}{$\ensuremath{\Gamma}(\ensuremath{\Omega}=2)$} & \multicolumn{1}{c}{Obs. $\ensuremath{\Omega}$}\\
      \midrule
      \endhead
      \bottomrule
      \multicolumn{11}{l}{Table continued on next page.}\\
      \endfoot
      \bottomrule
      \endlastfoot
      \footnotetext[1]{$\ensuremath{\Gamma}$ has units of \wnfwhm{} and all other parameters units of \wn{}. Fitting uncertainties are given parenthetically in units of the least significant digit and are relative only, not accounting for parameter correlation or any inadequacy in the spectral model specification. Fixed parameters are given without uncertainties. Blanked parameters were fixed to zero.}
      \footnotetext[2]{Term value $T$ is referenced to the \Xstate potential-energy minimum and is related to a virtual $T_{\ensuremath{\Sigma}=0,J=0}$ energy level according to $T_{\ensuremath{\Sigma}=0,J=0}=T+2B-2\ensuremath{\gamma}-\frac{4}{3}\ensuremath{\lambda}$.}
      \footnotetext[3]{Widths entered in this column are fitted to all $\ensuremath{\Omega}$ levels simultaneously, otherwise widths are given in the following columns for individual $\ensuremath{\Omega}$. A few bands have $J$ dependent widths as described in footnotes.}
      \footnotetext[4]{$\ensuremath{\Omega}$-states of this level that directly contribute to the observed absorption.}

  \\\multicolumn{11}{c}{\Large \ce{{}^{32}S {}^{16}O}}\\\\
  $X\,{}^3\Sigma^-(v=0)$\footnote{Computed from the isotopically-invariant ground-state parameters of \citet{lattanzi2015}. Additional parameters: $H=-2.16\times10^{-13}$, $\ensuremath{\lambda}_D=1.02\times10^{-5}$, $\ensuremath{\gamma}_D=-1.76\times10^{-8}$}  & $573.791\,05$     & $0.718$          & $1.13\times10^{-6}$       & \multicolumn{1}{c}{$-$}            & $5.28$       & $-0.005\,61$     & \multicolumn{1}{c}{$-$}                 & \multicolumn{1}{c}{$-$}                   & \multicolumn{1}{c}{$-$}                   & \multicolumn{1}{c}{$-$}                   & $0,1$      \\
\end{longtable}
  \end{small}
\end{landscape}

\begin{table*}
  \begin{minipage}{\linewidth}
    \caption{ Spin-orbit interaction energies, $\ensuremath{\xi}_{v_Bv_C}$.\protect\footnote{As defined in Eqs.~(\ref{eq:specific spin orbit matrix elements Omega1}) and (\ref{eq:specific spin orbit matrix elements Omega0}) and in units of \wn{}. Fitting uncertainties are given parenthetically in units of the least significant digit and are relative only, not accounting for parameter correlation or any inadequacy in the spectral model specification.}}
    \label{tab:interaction energies}
    \begin{center}
      \begin{tabular}{cllll}
        \toprule
        \multicolumn{1}{c}{Levels} & \multicolumn{1}{c}{\ce{{}^{32}S {}^{16}O}}  & \multicolumn{1}{c}{\ce{{}^{33}S {}^{16}O}}  & \multicolumn{1}{c}{\ce{{}^{34}S {}^{16}O}} & \multicolumn{1}{c}{\ce{{}^{36}S {}^{16}O}} \\
        \midrule
$B(5)/C(0)$          & $-8.86(13)$          & $-8.969(64)$         & $-9.030(69)$         & $-9.43(59)$         \\
$B(6)/C(1)$          & $-14.99(23)$         & $-14.74(20)$         & $-14$                & $-14.53(24)$        \\
$B(7)/C(2)$          & $-16.95(12)$         & $-16.59(32)$         & $-16.85(60)$         & $-17.80(62)$        \\
$B(8)/C(3)$          & $-14.88(28)$         & $-15.84(29)$         & $-15.39(51)$         & $-14.81(38)$        \\
$B(10)/C(4)$         & $-13.65(34)$         & \multicolumn{1}{c}{--} & $-14.86(19)$         & $-14.66(13)$        \\
$B(11)/C(5)$         & $-12.08(12)$         & $-11.92(22)$         & $-12.09(23)$         & $-12.41(19)$        \\
$B(12)/C(6)$         & $-10.7(12)$          & $-11.46(83)$         & $-10.60(78)$         & $-9.70(25)$         \\
$B(13)/C(7)$         & $-12.8(11)$          & $-10.9(12)$          & $-9.6(14)$           & $-8.08(72)$          \\
        \bottomrule
      \end{tabular}
    \end{center}
  \end{minipage}
\end{table*}

\begin{table*}
  \begin{minipage}{\linewidth}
    \caption{Deperturbed electric-dipole vibronic transition moments.\protect\footnote{In atomic units. Fitting uncertainties are given parenthetically in units of the least significant digit and are relative only, not accounting for parameter correlation or any inadequacy in the spectral model specification.}}
    \label{tab:dipole moments}
    \begin{center}
      \begin{tabular}{lllll}
        \toprule
        \multicolumn{1}{c}{Transition} & \multicolumn{1}{c}{\ce{{}^{32}S {}^{16}O}}  & \multicolumn{1}{c}{\ce{{}^{33}S {}^{16}O}}  & \multicolumn{1}{c}{\ce{{}^{34}S {}^{16}O}}  & \multicolumn{1}{c}{\ce{{}^{36}S {}^{16}O}} \\
        \midrule

$A(1)\leftarrow X(0)$ & $0.0179$             & \multicolumn{1}{c}{--} & \multicolumn{1}{c}{--} & \multicolumn{1}{c}{--}\\
$A(2)\leftarrow X(0)$ & $0.0235(20)$         & \multicolumn{1}{c}{--} & \multicolumn{1}{c}{--} & \multicolumn{1}{c}{--}\\
$A(3)\leftarrow X(0)$ & $0.0253(26)$         & \multicolumn{1}{c}{--} & \multicolumn{1}{c}{--} & \multicolumn{1}{c}{--}\\
$B(4)\leftarrow X(0)$ & $0.061(22)$          & $0.0469(42)$         & $0.0453(38)$         & $0.0543(43)$        \\
$B(5)\leftarrow X(0)$ & $0.0677(95)$         & $0.0656(47)$         & $0.0630(47)$         & $0.0751(47)$        \\
$B(6)\leftarrow X(0)$ & $0.0992(22)$         & $0.1044(20)$         & $0.0933(24)$         & $0.1046(13)$        \\
$B(7)\leftarrow X(0)$ & $0.1187(18)$         & $0.1034(29)$         & $0.1092(32)$         & $0.1250(13)$        \\
$B(8)\leftarrow X(0)$ & $0.1407(14)$         & $0.1370(23)$         & $0.1323(39)$         & $0.1476(10)$        \\
$B(9)\leftarrow X(0)$ & $0.16426(88)$        & $0.1560(17)$         & $0.1550(22)$         & $0.1650(12)$        \\
$B(10)\leftarrow X(0)$ & $0.1834(12)$         & $0.1773(25)$         & $0.1751(27)$         & $0.1803(13)$        \\
$B(11)\leftarrow X(0)$ & $0.1933(11)$         & $0.1946(28)$         & $0.1964(21)$         & $0.1998(12)$        \\
$B(12)\leftarrow X(0)$ & $0.2132(14)$         & $0.2108(35)$         & $0.2106(22)$         & $0.21154(89)$       \\
$B(13)\leftarrow X(0)$ & $0.2188(26)$         & $0.2182(40)$         & $0.2161(35)$         & $0.2166(16)$        \\
$B(14)\leftarrow X(0)$ & $0.2168(22)$         & $0.2185(34)$         & $0.2211(23)$         & $0.21005(80)$       \\
$B(15)\leftarrow X(0)$ & $0.21897(86)$        & $0.2224(33)$         & $0.2257(21)$         & $0.21618(78)$       \\
$B(16)\leftarrow X(0)$ & $0.20123(72)$        & $0.2079(48)$         & $0.2117(31)$         & $0.2026(10)$        \\
$B(17)\leftarrow X(0)$ & $0.1621(10)$         & $0.1773(96)$         & $0.1768(62)$         & $0.1749(20)$        \\
$B(18)\leftarrow X(0)$ & $0.1424(11)$         & \multicolumn{1}{c}{--} & \multicolumn{1}{c}{--} & $0.1441(32)$        \\
$B(19)\leftarrow X(0)$ & $0.14232(94)$        & \multicolumn{1}{c}{--} & \multicolumn{1}{c}{--} & $0.1533(39)$        \\
$B(20)\leftarrow X(0)$ & $0.15346(96)$        & \multicolumn{1}{c}{--} & \multicolumn{1}{c}{--} & $0.1243(64)$        \\
$B(21)\leftarrow X(0)$ & $0.13924(79)$        & \multicolumn{1}{c}{--} & \multicolumn{1}{c}{--} & \multicolumn{1}{c}{--}\\
$B(22)\leftarrow X(0)$ & $0.1344(10)$         & \multicolumn{1}{c}{--} & \multicolumn{1}{c}{--} & \multicolumn{1}{c}{--}\\
$B(23)\leftarrow X(0)$ & $0.1199(10)$         & \multicolumn{1}{c}{--} & \multicolumn{1}{c}{--} & \multicolumn{1}{c}{--}\\
$B(24)\leftarrow X(0)$ & $0.1174(12)$         & \multicolumn{1}{c}{--} & \multicolumn{1}{c}{--} & \multicolumn{1}{c}{--}\\
$B(25)\leftarrow X(0)$ & $0.1178(12)$         & \multicolumn{1}{c}{--} & \multicolumn{1}{c}{--} & \multicolumn{1}{c}{--}\\
$B(26)\leftarrow X(0)$ & $0.100$              & \multicolumn{1}{c}{--} & \multicolumn{1}{c}{--} & \multicolumn{1}{c}{--}\\
$B(27)\leftarrow X(0)$ & $0.0931$             & \multicolumn{1}{c}{--} & \multicolumn{1}{c}{--} & \multicolumn{1}{c}{--}\\
$B(28)\leftarrow X(0)$ & $0.0857$             & \multicolumn{1}{c}{--} & \multicolumn{1}{c}{--} & \multicolumn{1}{c}{--}\\
$B(29)\leftarrow X(0)$ & $0.0783$             & \multicolumn{1}{c}{--} & \multicolumn{1}{c}{--} & \multicolumn{1}{c}{--}\\
$B(30)\leftarrow X(0)$ & $0.0705$             & \multicolumn{1}{c}{--} & \multicolumn{1}{c}{--} & \multicolumn{1}{c}{--}\\
$C(1)\leftarrow X(0)$ & $0.0160(42)$         & \multicolumn{1}{c}{--} & \multicolumn{1}{c}{--} & \multicolumn{1}{c}{--}\\
$C(2)\leftarrow X(0)$ & $0.0315(31)$         & $0.0542(26)$         & $0.0402(39)$         & $0.0316(21)$        \\
$C(3)\leftarrow X(0)$ & $0.0403(18)$         & $0.0292(38)$         & $0.0323(47)$         & $0.0476(18)$        \\
$C(4)\leftarrow X(0)$ & $0.0523(16)$         & \multicolumn{1}{c}{--} & $0.0431(40)$         & $0.0509(20)$        \\
$C(5)\leftarrow X(0)$ & $0.0386(23)$         & $0.0483(46)$         & $0.0289(54)$         & $0.0503(26)$        \\
$C(6)\leftarrow X(0)$ & $0.0417(38)$         & $0.0446(74)$         & $0.0478(49)$         & $0.0441(24)$        \\
$C(7)\leftarrow X(0)$ & $0.0327(60)$         & $0.033(11)$          & $0.0482(48)$         & $0.0315(39)$        \\

      \end{tabular}
    \end{center}
  \end{minipage}
\end{table*}

\section{Results}
\label{sec:results}

A detailed discussion of the observed SO bands is provided in this section.
Listings of the deduced deperturbed molecular constants of $\Astate(v)$, $\Bstate(v)$, and $\Cstate(v)$ electronic-vibrational levels and their predissociation widths, along with $\Bstate(v_B)\sim\Cstate(v_C)$ interaction parameters, and \AX{v=1-3}, \BX{v=4-30} and \CX{v=0-7} transition moments are given in Tables~\ref{tab:level parameters}, \ref{tab:interaction energies}, and \ref{tab:dipole moments}, respectively.
In order to provide data directly comparable with the experimental spectra, a full list of perturbed (coupled) level energies and predissociation widths, as well as line frequencies, widths, and intensities is provided in an online appendix \citep{online_appendix}.
All level energies are given relative to the ground-state equilibrium energy with $X(v=0)$ vibrational energies ($T$ in Table~\ref{tab:level parameters}) computed from the isotopically-invariant parameters of \citet{lattanzi2015}: 573.79, 570.89, 568.16, and \wn{563.09} for \ce{{}^{32}S {}^{16}O}, \ce{{}^{33}S {}^{16}O}, \ce{{}^{34}S {}^{16}O}, and \ce{{}^{36}S {}^{16}O}, respectively.
Spin and rotational constants for $\Xstate(v=0)$ are also taken from \citet{lattanzi2015}.

\subsection{\boldmath$B(4)\leftarrow X(0)$}

The predissociation-broadened \BX{4} spectrum is plotted in Fig.~\ref{fig:spectrum B04-X00} and is overlapped with significant absorption from $B(6)\leftarrow{}X(1)$.
This is highlighted in Fig.~\ref{fig:spectrum B04-X00} by the residual error of a model neglecting this extra absorption.

\subsection{\boldmath$B(5)/C(0) \leftarrow X(0)$}
\label{sec:B(5)/C(0)}

Previous analyses of \ce{{}^{32}S {}^{16}O} \BX{5} and \CX{0} photoabsorption \citep{liu_ching-ping2006} and multiphoton-ionisation \citep{archer2000} spectra reveal spin-orbit and rotational interactions locally mixing the $B(5)$ and $C(0)$ states.
Further interaction of $C(0)$ with $d(1)$ and additional $\Lambda$-doubling of $C(0)$ are also revealed by these studies.
We observed the \BX{5} transition in four isotopologues and find similar interactions occurring between $B(5)$ and $C(0)$ in all of them.

An experimental \ce{{}^{32}S {}^{16}O} \BX{5} absorption spectrum is shown in Fig.~\ref{fig:spectrum B05-X00} and is overlapped with \ce{SO2} absorption.
The signal attributable to SO after accounting for \ce{SO2} and hot-band contamination is plotted as a residual error of a model neglecting SO absorption from $X(0)$.
Our measurements are significantly less sensitive than the laser-based absorption of \citet{liu_ching-ping2006} but our analysis benefits from a well defined rotational temperature and predictable line intensities.

No \CX{0} or \dX{1} absorption is evident in our spectra and no sensitivity to the $C(0)\sim d(1)$ interaction was found.
We then neglect $d(1)$ in our analysis of \ce{{}^{33}S {}^{16}O},  \ce{{}^{34}S {}^{16}O} and \ce{{}^{36}S {}^{16}O}, and, for self-consistency, also for \ce{{}^{32}S {}^{16}O}.
The perturbative influence of $C(v=0,\ensuremath{\Omega}=0)$ on $B(5)$ is significant in all isotopologues and the $C(0)$ level was included along with a spin-orbit interaction parameter.
The $C(0)$ spin-orbit and spin-spin constants were fixed to values in line with our measurements of other \CX{v} bands.
Our \ce{{}^{32}S {}^{16}O} $B(5)\sim C(0)$ model differs from that of \citet{liu_ching-ping2006} and will not therefore reproduce their spectrum of $C(0)$ $\ensuremath{\Omega}=2$ and 1 levels that are perturbed by $d(1)$.

\subsection{\boldmath$B(6)/C(1) \leftarrow X(0)$}

The $B(6)$ level is perturbed by $C(1)$ and there is a level crossing of their nominal $\ensuremath{\Sigma}=0$ levels near $J=30$.
Direct \CX{1} absorption is only observed for \ce{{}^{32}S {}^{16}O} but its interaction with $B(6)$ is strong enough to constrain some molecular parameters in all isotopologues and a spin-orbit interaction energy in the cases of \ce{{}^{33}S {}^{16}O} and \ce{{}^{36}S {}^{16}O}.
The broadening of $C(1)$ levels deduced from its appearance in \ce{{}^{32}S {}^{16}O} is near to or below the sensitivity of the experiment and we estimate a rigorous upper limit of \wnfwhm{0.1}.

 \subsection{\boldmath$B(7)/C(2)\leftarrow X(0)$}
\begin{figure*}
  \centering
  \includegraphics[width=\columnwidth]{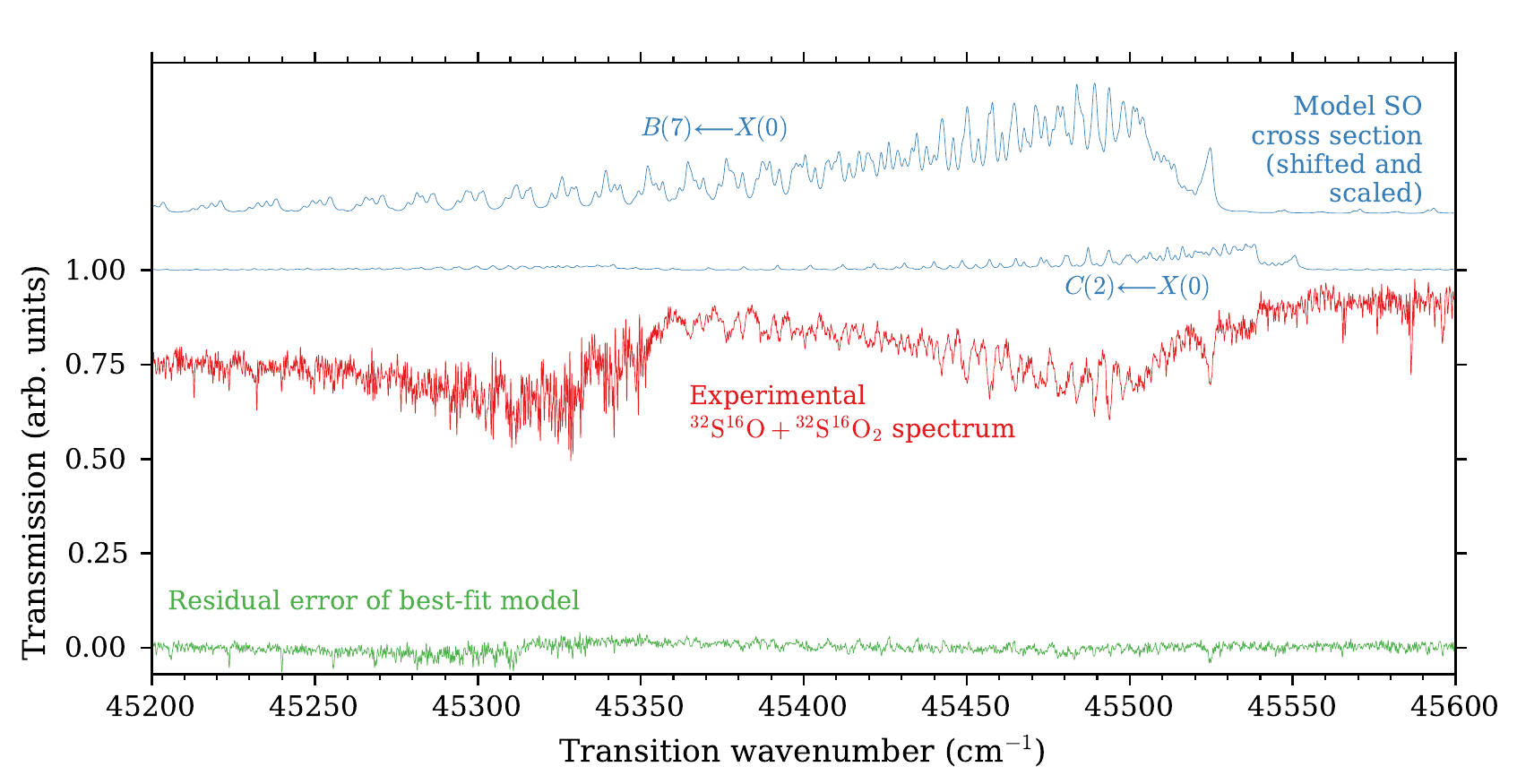}
  \caption{An experimental photoabsorption spectrum showing \ce{{}^{32}S {}^{16}O} \BX{7} and \CX{2} and the residual error of a best-fit model. The model SO cross section is also indicated with separate contributions from nominal \BX{7} and \CX{2} transitions.}
  \label{fig:spectrum B07-X00}
\end{figure*}
This band is visibly perturbed and has an unusual multi-headed band structure that is most obvious for  \ce{{}^{32}S {}^{16}O}.
This is shown in Fig.~\ref{fig:spectrum B07-X00} and is due to an interaction between $B(7)$ and the $\Omega=1$ level of $C(2)$, as discussed previously \citep{ornellas1998b,liu_ching-ping2006}.
 A best-fit model of our measured spectra places the $C(v=2,\ensuremath{\Omega}=1)$ level at slightly higher energy than $B(7)$, with no crossing of their rotational series, and includes significant absorption from \CX{2} transitions.
 A similar picture applies to \ce{{}^{33}S {}^{16}O}, \ce{{}^{34}S {}^{16}O}, and \ce{{}^{36}S {}^{16}O} absorption but these are less dramatically perturbed because of a greater separation of mass-shifted $B(7)$ and $C(2)$ levels.  

\subsection{\boldmath${B(8)/C(3) \leftarrow X(0)}$}
\label{sec:B(8)/C(3)}

\begin{figure*}
  \centering
  \includegraphics[width=\columnwidth]{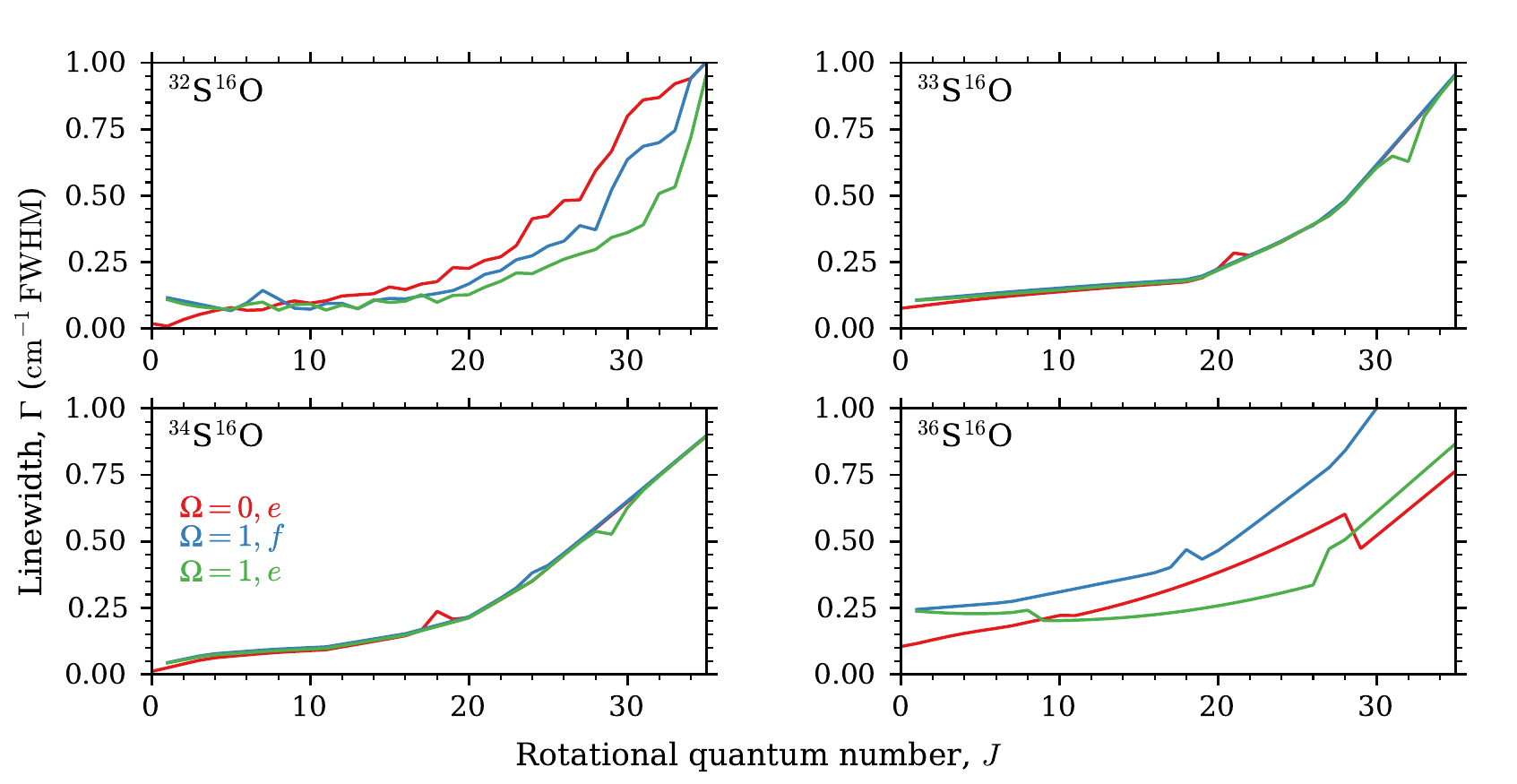}
  \caption{Perturbed linewidths fitted to $B(8)$.}
  \label{fig:spectrum B08 widths}
\end{figure*}

This is the least predissociation-broadened \BX{v} band appearing in our spectra, and its rotational structure is analysed in greater detail.
We fitted multiple absorption spectra showing \BX{8} for each isotopologue and observe \CX{3} absorption, perturbations mixing $B(8)$ and $C(3)$, and $J$- and $\ensuremath{\Omega}$-dependent predissociation widths.
A measured and modelled spectrum of \ce{{}^{32}S {}^{16}O} \BX{8} and \CX{3} is shown in Fig.~\ref{fig:spectrum B08-X00} as well as a reference \ce{{}^{32}S {}^{16}O_2} spectrum used to account for contamination.

Trial models assuming deperturbed $B(8)$ $J$- and $\ensuremath{\Omega}$-independent linewidths imperfectly fitted measured spectra of all isotopologues.
More freely fitted widths are shown in Fig.~\ref{fig:spectrum B08 widths} and satisfactorily fitted all data.

The deperturbed $B(8)$ widths of \ce{{}^{33}S {}^{16}O} and \ce{{}^{34}S {}^{16}O} were modelled as $\ensuremath{\Omega}$-independent and piecewise-linear in $J$.
The perturbed widths are shown in Fig.~\ref{fig:spectrum B08 widths} and show local perturbations where the $B(8)$ sublevels are crossed by $C(v=3,\ensuremath{\Omega}=1)$.
The spectra showing \ce{{}^{32}S {}^{16}O} are of sufficient quality that the perturbed widths of all $B(8)$ levels for $6<J<33$ are fitted individually, as shown in Fig.~\ref{fig:spectrum B08 widths}, and show increasingly broad spin-sublevels in the ordering $\ensuremath{\Gamma}_{\ensuremath{\Omega}=1,e}<\ensuremath{\Gamma}_{\ensuremath{\Omega}=1,f}<\ensuremath{\Gamma}_{\ensuremath{\Omega}=0,e}$.
A different width ordering was found in the case of \ce{{}^{36}S {}^{16}O} for levels with $J>10$ with $\ensuremath{\Gamma}_{\ensuremath{\Omega}=1,e}<\ensuremath{\Gamma}_{\ensuremath{\Omega}=0,e}<\ensuremath{\Gamma}_{\ensuremath{\Omega}=1,f}$.
The \ce{{}^{36}S {}^{16}O} widths shown in Fig.~\ref{fig:spectrum B08 widths} are less constrained by the experimental data but their overall $\ensuremath{\Omega}$ and $J$ dependencies are necessary to reproduce our spectra.
The possibility of similar-magnitude $\ensuremath{\Omega}$-dependences for the \ce{{}^{33}S {}^{16}O} and \ce{{}^{34}S {}^{16}O} widths is not ruled out.

Modelled $J$-independent and $\ensuremath{\Omega}$-dependent deperturbed widths were assumed for $C(3)$, with values for the $\ensuremath{\Omega}=0$, 1, and 2 \ce{{}^{32}S {}^{16}O} levels found to be $0.20\pm{}0.05$, $1.4\pm{}0.4$, and $0.4\pm{}0.1$\wnfwhm{}, respectively.
This strong $\ensuremath{\Omega}$-dependence is evident in Fig.~\ref{fig:spectrum B08-X00} where the \CX{3} transition moment has been set to zero to reveal its absorption as a model residual error.
Spectra containing \ce{{}^{33}S {}^{16}O} and \ce{{}^{34}S {}^{16}O} absorption are relatively noisy and the $C(3)$ widths similar to \ce{{}^{32}S {}^{16}O} were assumed for these isotopologues.
A different $\ensuremath{\Omega}$-ordering of $C(3)$ linewidths was tentatively deduced from the \ce{{}^{36}S {}^{16}O} spectrum with $0.8\pm{}0.4$, $1.3\pm{}0.5$, and $0.6\pm{}0.2$\wnfwhm{} for $\ensuremath{\Omega}=0$, 1, and 2, respectively.

\subsection{\boldmath$B(9) \leftarrow X(0)$}

No interaction between $B(9)$ and any $C(v)$ level was required to adequately reproduce the observed \BX{9} absorption (shown in Fig.~\ref{fig:spectrum B09-X00} for \ce{{}^{32}S {}^{16}O}) and any actual interaction with the neighbouring but non-crossing $C(3)$ and $C(4)$ levels has been adequately incorporated into the molecular parameters of $B(9)$.
It was necessary to assume quite different predissociation broadening for the $\ensuremath{\Omega}=0$ and 1 sublevels, with consistent values of about 0.6 and \wnfwhm{3}, respectively, found for all four isotopologues.

\subsection{\boldmath$B(10)/C(4) \leftarrow X(0)$}
A \ce{{}^{32}S {}^{16}O} absorption spectrum of the region containing \BX{10}, and \CX{4} is shown in Fig.~\ref{fig:spectrum B10-X00} along with the absorption due to SO of all isotopologues highlighted by models accounting for all spectral contributions apart from SO.
In this figure, rotational structure arising from all three $\ensuremath{\Omega}$-components of $C(4)$ in \ce{{}^{32}S {}^{16}O} is clearly resolved and is well-modelled along with its spin-orbit interaction with $B(10)$.
This model also necessitated the inclusion of $C(4)$ $\ensuremath{\Lambda}$-doubling parameters to best fit the \ce{{}^{32}S {}^{16}O} spectra.
An attempt to replace the fitted $\ensuremath{\Lambda}$-doubling parameters with an additional rotational interaction between $B(10)$ and $C(4)$ was ineffective, as were trial additions of $B(9)/C(4)$ and $B(11)/C(4)$ interactions.

Too few \CX{4} lines appear in our spectra of \ce{{}^{33}S {}^{16}O} for a positive assignment to be made.
A list of unassigned lines comprising \CX{v=4,\ensuremath{\Omega}=0} is given in the online appendix.
Fewer \CX{4} lines are evident in our \ce{{}^{34}S {}^{16}O} spectrum than for \ce{{}^{32}S {}^{16}O} so the deperturbed width of $C(4)$ $\ensuremath{\Omega}=1$ levels was fixed to the \ce{{}^{32}S {}^{16}O} value.

\subsection{\boldmath$B(11)/C(5)  \leftarrow X(0)$}
The $B(11)$ level is perturbed by $C(5)$, which itself contributes significantly to the observed \ce{{}^{32}S {}^{16}O} spectrum, as shown in Fig.~\ref{fig:spectrum B11-X00}.
Only a weak signal of \CX{v=5,\ensuremath{\Omega}=0\text{ and }1} absorption is evident in our \ce{{}^{34}S {}^{16}O} spectra so these levels have widths fixed to their \ce{{}^{32}S {}^{16}O} values, while the $\ensuremath{\Omega}$-dependent widths of \ce{{}^{33}S {}^{16}O} are marginally measurable.

\subsection{\boldmath$B(12)/C(6)  \leftarrow X(0)$}
There is a clear spin-orbit interaction mixing $B(12)$ and $C(6)$, with weak \CX{6} absorption evident for all isotopologues.

The measured predissociation widths of \ce{{}^{32}S {}^{16}O} and \ce{{}^{36}S {}^{16}O} $C(v=6,\ensuremath{\Omega}=0)$ levels are significantly broader than for either $\ensuremath{\Omega}=1$ or 2,  with fitted $\ensuremath{\Omega}=0$ widths of $6.5\pm{}0.3$ and $2\pm{}1$\,\wnfwhm{}, respectively.
It is difficult to estimate the true uncertainty of these widths from the weakly-absorbing \CX{6} transitions, but they have clear lower bounds of 2 and \wnfwhm{1.5}, respectively.
$C(v=6,\ensuremath{\Omega}=0)$ widths could not be measured for \ce{{}^{33}S {}^{16}O} and \ce{{}^{34}S {}^{16}O}.

\subsection{\boldmath$B(13)/C(7) \leftarrow X(0)$}
A spin-orbit interaction between $B(13)$ and $C(7)$ was determined for all isotopologues and constrained by weak \CX{7} absorption that is too broadened to reveal any distinct line structure.
The deperturbed $\ensuremath{\Omega}=1$ linewidths of $B(13)$ are consistently smaller than for $\ensuremath{\Omega}=0$ in the \ce{{}^{32}S {}^{16}O}, \ce{{}^{34}S {}^{16}O}, and \ce{{}^{36}S {}^{16}O} isotopologues, and a similar $\ensuremath{\Omega}=0$ width was assumed for \ce{{}^{33}S {}^{16}O}.

\subsection{\boldmath$B(14) \leftarrow X(0)$}
The $B(14)$ level is more broadened than most other $B(v)$ vibrational levels, and only term origins, rotational constants and $J$- and $\ensuremath{\Omega}$-independent widths are fitted to the observed \BX{14} spectrum.

\subsection{\boldmath$B(15) \leftarrow X(0)$}
\label{sec:B15}

\begin{figure*}
  \centering
  \includegraphics[width=\columnwidth]{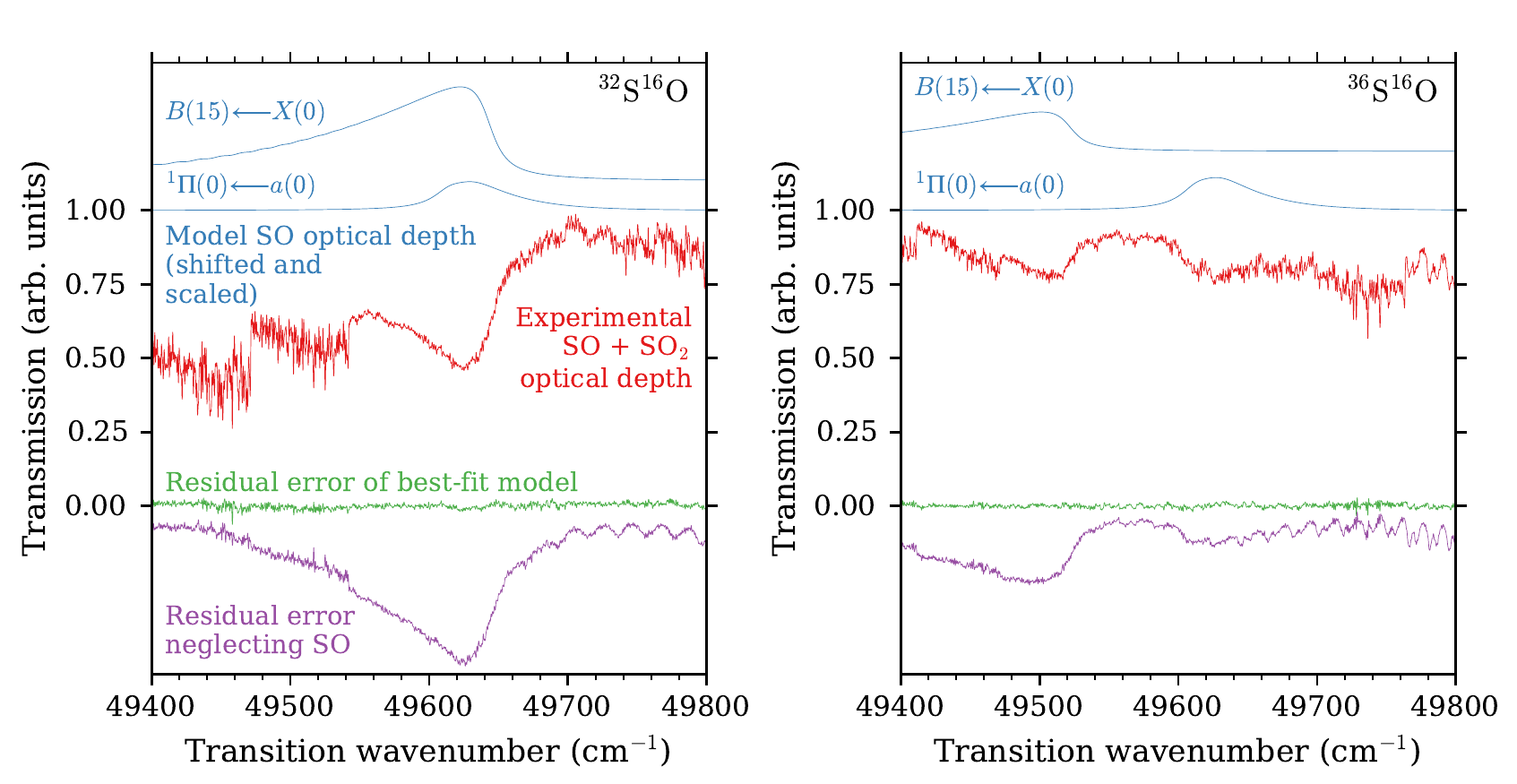}
  \caption{Experimental and modelled spectra of \BX{15} and overlapping SO and \ce{SO2} absorption in two isotopologues.  The residual error neglecting SO includes some contribution from high-excitation rotational transitions of \BX{16}.}
  \label{fig:spectrum B15-X00}
\end{figure*} 

The analysis of \BX{15} was complicated by its overlap with a broadened and previously unobserved SO absorption band that occurs near \wn{49620} in all isotopologues.
We assign this to absorption from the metastable $\astate(v=0)$ level non-thermally excited in the discharge to a previously-unobserved ${}^1\ensuremath{\Pi}(v=0)$ upper level.
A further difficulty arises because $B(15)$ occurs at an energy near the configurational change of the \Bstate state.
Its rotational constant cannot then be reliably extrapolated from lower vibrational levels and is a key constraint on the irregular shape of the \Bstate potential-energy curve near 2.4\,\AA, as shown in Fig.~\ref{fig:introduction PECs}.

To disentangle $B-X$ and $\ensuremath{{}^1\Pi}-a$ absorption we simultaneously analysed all isotopologues and simulated the new \ensuremath{{}^1\Pi} level with Dunham constants for \ce{{}^{32}S {}^{16}O} and mass-scaled these \citep{le_roy1999} to the other isotopologues.
The relevant spectra and band-simulations for \ce{{}^{32}S {}^{16}O} and \ce{{}^{36}S {}^{16}O} are shown in Fig.~\ref{fig:spectrum B15-X00}, with the two bands completely overlapped in the former case and well separated but less prominent in the latter spectrum.
Intermediate overlap occurs for \ce{{}^{33}S {}^{16}O} and \ce{{}^{34}S {}^{16}O}.
Mass-independent line broadening was assumed for both the \ensuremath{{}^1\Pi} level and $B(15)$.

We defer a detailed discussion of absorption observed in our spectra originating from $\astate$ to a future study, but the proposed upper state assignment of the \wn{49620} band to a $\ensuremath{{}^1\Pi}(v=0)$ fundamental level is based on the electronic state constants determined under this assumption. These are $T_e=\np{55397}\pm5$ and $\ensuremath{\omega}_e=1321\pm5$, $B_e={0.766\pm0.010}$, and $\ensuremath{\alpha}_e=\-0.025\pm0.020$\,cm$^{-1}$ for \ce{{}^{32}S {}^{16}O}.
The separate determinations of $T_e$ and $\ensuremath{\omega}_e$, and $B_e$ and $\ensuremath{\alpha}_e$ is possible because of the identification of additional new absorption attributed to the corresponding $\ensuremath{{}^1\Pi}(v=1)$ level overlapping \BX{20}, and  discussed in Sec.~\ref{sec:B(20)}, and supported by the range of measured isotopologues.

\subsection{\boldmath$B(16) \leftarrow X(0)$}
The $B(16)$ level is less predissociated than the neighbouring vibrational levels and the fitted widths decrease with increasing reduced mass.
The width fitted to deperturbed $\ensuremath{\Omega}=0$ levels is slightly greater than for $\ensuremath{\Omega}=1$ in \ce{{}^{32}S {}^{16}O} but no such distinction was observed in the other isotopologues despite their being observed with similar precision.
An anomalously large $B(16)$ centrifugal distortion was fitted to the \ce{{}^{32}S {}^{16}O} spectrum but not for the other isotopologues.

\subsection{\boldmath$B(17-30) \leftarrow X(0)$}
\label{sec:B(20)}

\begin{figure*}
  \centering
  \includegraphics[width=\columnwidth]{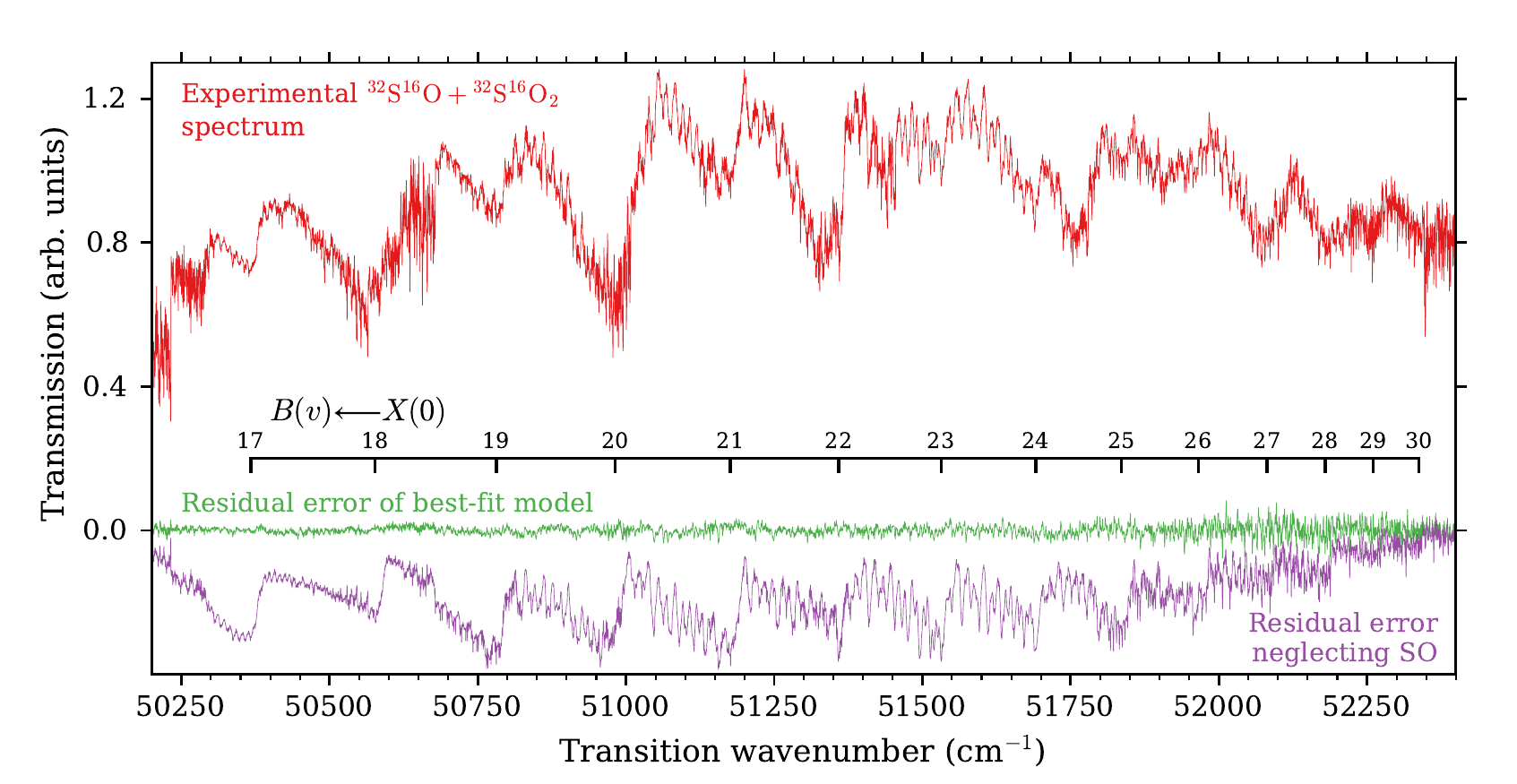}
  \caption{Measured spectrum of \ce{{}^{32}S {}^{16}O} showing \BX{25-30}, as well as the residual error of a best-fit model and neglecting \ce{{}^{32}S {}^{16}O} absorption.\label{fig:B25-X00_to_B30-X00_spectrum}}
\end{figure*}

The $\BX{v}$ progression was followed in all isotopologues as far as $v=17$ with additional measurements approaching the \Bstate dissociation limit made for \ce{{}^{32}S {}^{16}O}, with an example spectrum shown in Fig.~\ref{fig:B25-X00_to_B30-X00_spectrum}, and as far as $v=20$ in \ce{{}^{36}S {}^{16}O}.
The decreasing vibrational spacing of $B(v)$ levels leads to severely overlapped  rotational structure near the dissociation limit.

Widths were fitted to the \BX{v=17-25} levels independently of $\ensuremath{\Omega}$.
The fitted parameters governing \BX{v=26-30} absorption are less satisfactory due to the weakness of these bands, decreasing linewidths approaching the dissociation limit, and a complete overlap of their rotational structure.
Reasonable agreement with the resolved rotational structure in the region of \BX{28} was found assuming a fixed broadening of \wnfwhm{1} for $B(28)$, but no consistent fit to the many overlapping lines at higher frequencies could be found.
These are then represented approximately by absorption into effective $B(29)$ and $B(30)$ levels artificially broadened by \wnfwhm{15} to smooth over their confused band profiles.
Additionally, it was necessary to constrain the transition moments of \BX{26-30} absorption bands to values extrapolated from lower vibrational levels.
This extrapolation was made assuming an $R$-independent electronic transition moment, as presented in Sec.~\ref{sec:transition moments}.

\subsection{\boldmath$A(1-3)\leftarrow X(0)$}
\label{sec:A-X}

\begin{figure*}
  \centering
  \includegraphics[width=\columnwidth]{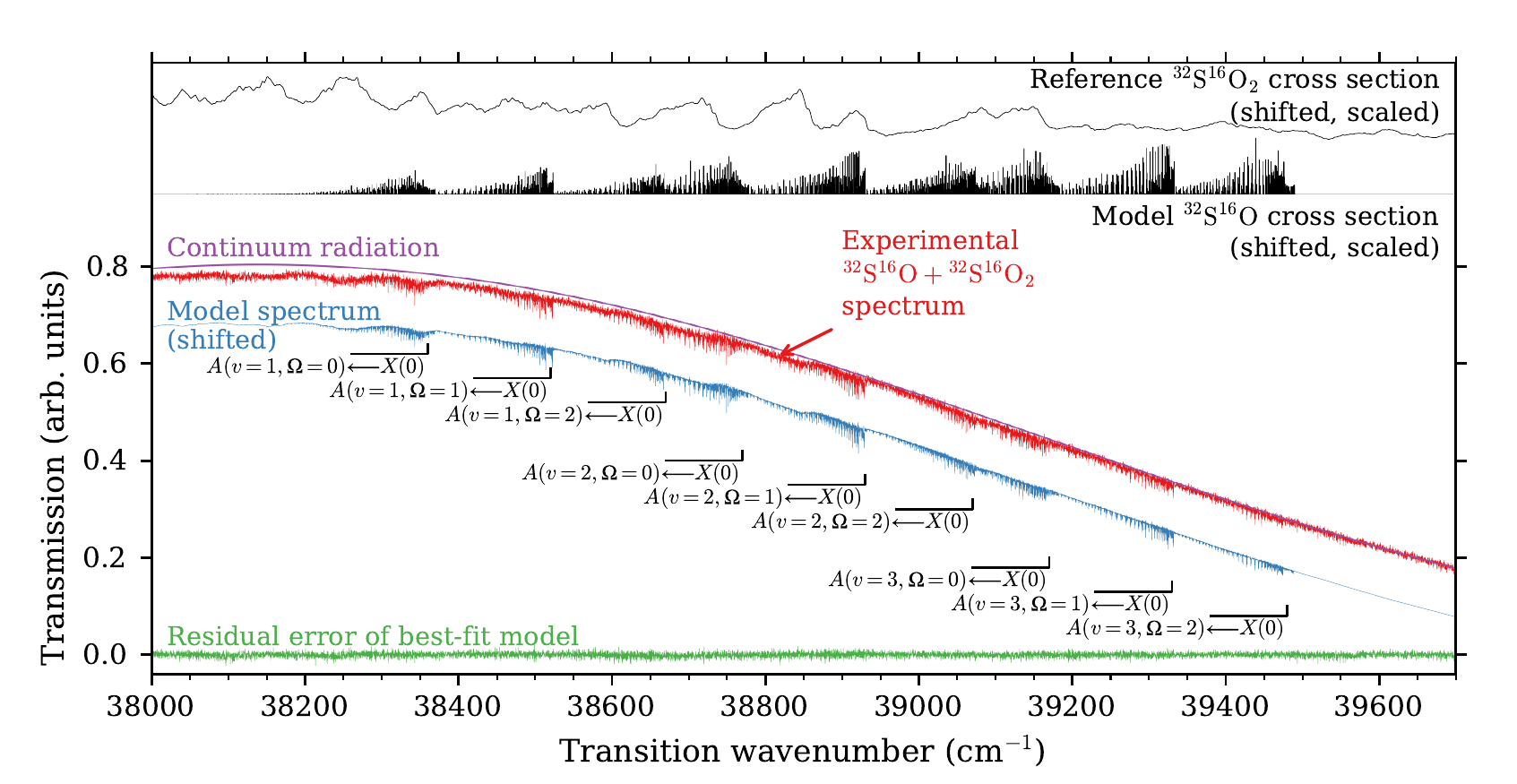}
  \caption{An experimental spectrum showing \ce{{}^{32}S {}^{16}O} \AX{1}, \AX{2}, and \AX{3} absorption compared with a model simulation accounting for variable background continuum radiation and absorption by \ce{{}^{32}S {}^{16}O} and \ce{{}^{32}S {}^{16}O_2}, and the residual error of this model.}
  \label{fig:A-X spectrum}
\end{figure*}
 
The \AX{v=1-3} absorption bands are weakly observed between \np{38000} and \myinvcm{39500} in the \ce{{}^{32}S {}^{16}O_2} discharge, and their rotational structure is illustrated in Fig.~\ref{fig:A-X spectrum} along with an effective-Hamiltonian band model.
These were measured and analysed in order to calibrate the column density of SO radicals.

The $A(1)$ level was modelled with the molecular parameters deduced by \citet{elks1999} with consideration of a local perturbation shifting its $\ensuremath{\Omega}=1$ $J=15$ level \citep{colin1982}.
Entirely new parameterisations were made for the $A(2)$ and $A(3)$ levels and fitted the experimental spectrum marginally better than the $A(2)$ constants of \citet{elks1999} and significantly better in the case of $A(3)$, for which previous measurements are relatively limited.

The ratio of squared transition moments fitted to the \AX{v} bands with $v=1$:2:3 is 1:1.71:1.63 and is comparable with the ratio of Einstein-$A$ coefficients computed for these transitions by \citet{borin2000}, 1:1.47:1.61.
The difference in $A(2)/A(1)$ ratios may result from a greater error in our fitted transition moments than implied by their statistical uncertainties listed in Table~\ref{tab:level parameters}, although a manual adjustment to match the \citet{borin2000} ratio visibly degrades the experimental fits.
Alternatively, there may be some error associated with the highly $R$-dependent electronic transition moment controlling the calculated ratios \citep{elks1999,borin2000}.

The total decay rate of $A(1)$ by all processes is experimentally known \citep{elks1999} and is $J$- and $\Omega$-independent with a measured value of $A^\text{total}_{A(1)}=\np{74100}\pm1100$\,s$^{-1}$.
The relative and rotationally-unresolved branching ratios, $\ensuremath{\eta}_{A(1)\rightarrow X(v'')}$, for partial decay via emission to  $v''=0$ to 11 are also measured \citep{lo1988}.
We deduce a band-averaged emission rate, $A_{A(1)\rightarrow X(0)}$, from these data according to:
\begin{align}
  A_{A(1)\rightarrow X(0)} &= A^\text{total}_{A(1)} \frac{\ensuremath{\eta}_{A(1)\rightarrow X(0)}}{\sum_{v=0}^{11} \ensuremath{\eta}_{A(1)\rightarrow X(v)}}  \notag \\
                           &= (3.73\pm 0.37)\times 10^5\,\text{s}^{-1},
  \label{eq:A-X(1,0) partial lifetime}
\end{align}
and also a band-integrated absorption $f$-value and electronic-vibrational transition moment:
\begin{align}
  \label{eq:A-X(1,0) fvalue}
  f_{\AX{1}} &= (7.55\pm 0.76)\times 10^{-5}, \\
  \text{and }\ensuremath{\mu}_{\AX{1}} &= 0.0179\pm0.0009\,\text{au},
\end{align}
respectively, using Eqs.~(\ref{eq:band fvalue}) and (\ref{eq:band emission rate}).
We assign a 10\% uncertainty to $A_{A(1)\rightarrow X(0)}$ and $f_{\AX{1}}$, corresponding to a 5\% uncertainty in $\ensuremath{\mu}_{\AX{1}}$ that is higher than estimated for the $A(1)$ lifetime by \citet{elks1999}.
This is to account for possible systematic error in the experimental lifetime, which is 30\% larger than an earlier determination \citep{clyne1982}, and a possible error contribution from the experimental emission branching ratios.

A trial calculation was made of $A(1)\rightarrow X(v'')$ emission rates using \Xstate and \Astate potential-energy curves taken from \citet{lattanzi2015} and \citet{sarka2019b}, respectively, and the experimentally-deduced $\ensuremath{\mu}_{A-X}$ electronic transition moment of \citet{elks1999}.
This confirmed that emission to levels with $v''>11$ is negligible
and the experimental branching ratios in Eq.~(\ref{eq:A-X(1,0) partial lifetime}) are adequately normalised when compared with their statistical uncertainties.

By fixing our model \AX{1} transition moment using the $f$-value from Eq.~(\ref{eq:A-X(1,0) fvalue}) we constrain the \ce{{}^{32}S {}^{16}O} column-density in Fig.~\ref{fig:A-X spectrum} to $(2.0\pm0.4)\times 10^{16}$\,cm$^{-2}$, where the fractional uncertainty is greater than for the reference $f$-value due to  the fitting uncertainty of our rather weak \AX{1} spectrum.
The separation in frequency between the \AX{v} bands, and \BX{v} and \CX{v} band is too great to permit their simultaneous measurement at DESIRS.
Instead, the \AX{1} and \BX{8} bands were measured consecutively under identical discharge conditions and a calibration of all \ce{{}^{32}S {}^{16}O}-spectra column densities thus attained.
This procedure was repeated several times during the experiment and the ratio of \AX{v} to \BX{v} $f$-values and found to be consistent within 5\%.

A rotational temperature of 290\,K  was found to best match $A-X$ spectra and is somewhat lower than the 360\,K temperature consistently deduced from the $B-X$ spectrum.
The reason for this difference is not clear but comparing fitted spectra of $A-X$ at 290 and 360\,K only results in a marginal change in their quality-of-fit and $f$-values that differ by about 5\%.
Further confirmation of the SO column density deduced here is provided in Secs.~\ref{sec:transition moments} and \ref{sec:comparison with previous cross sections}. 

\section{Discussion}
\label{sec:discussion}

\begin{figure}
  \centering
  \includegraphics{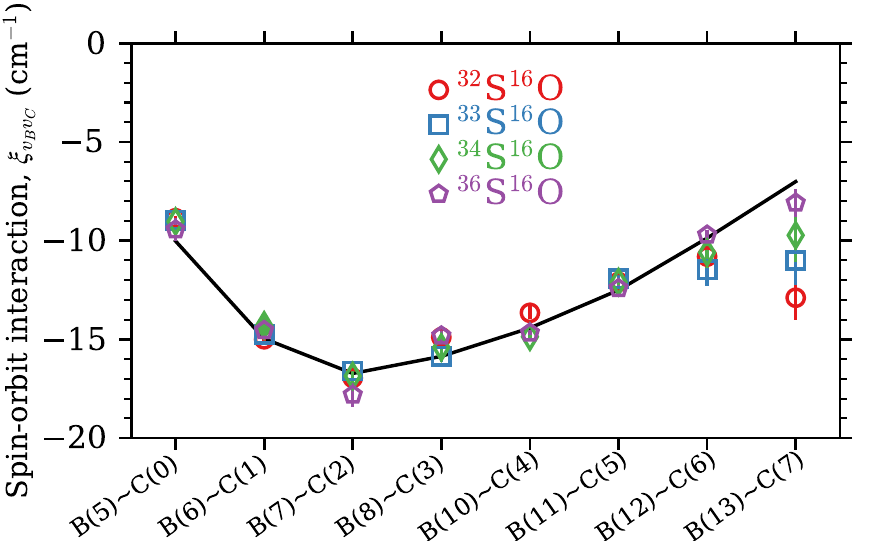}
  \caption{Spin-orbit interaction energies of neighbouring $B(v_B)$ and $C(v_C)$ states.  \emph{Symbols:} Band-by-band fitted parameters. \emph{Curve vertices:} \protect\ce{{}^{32}S {}^{16}O} interaction energies computed from the global model.}
  \label{fig:interactions}
\end{figure}

\begin{figure}
  \centering
  \includegraphics{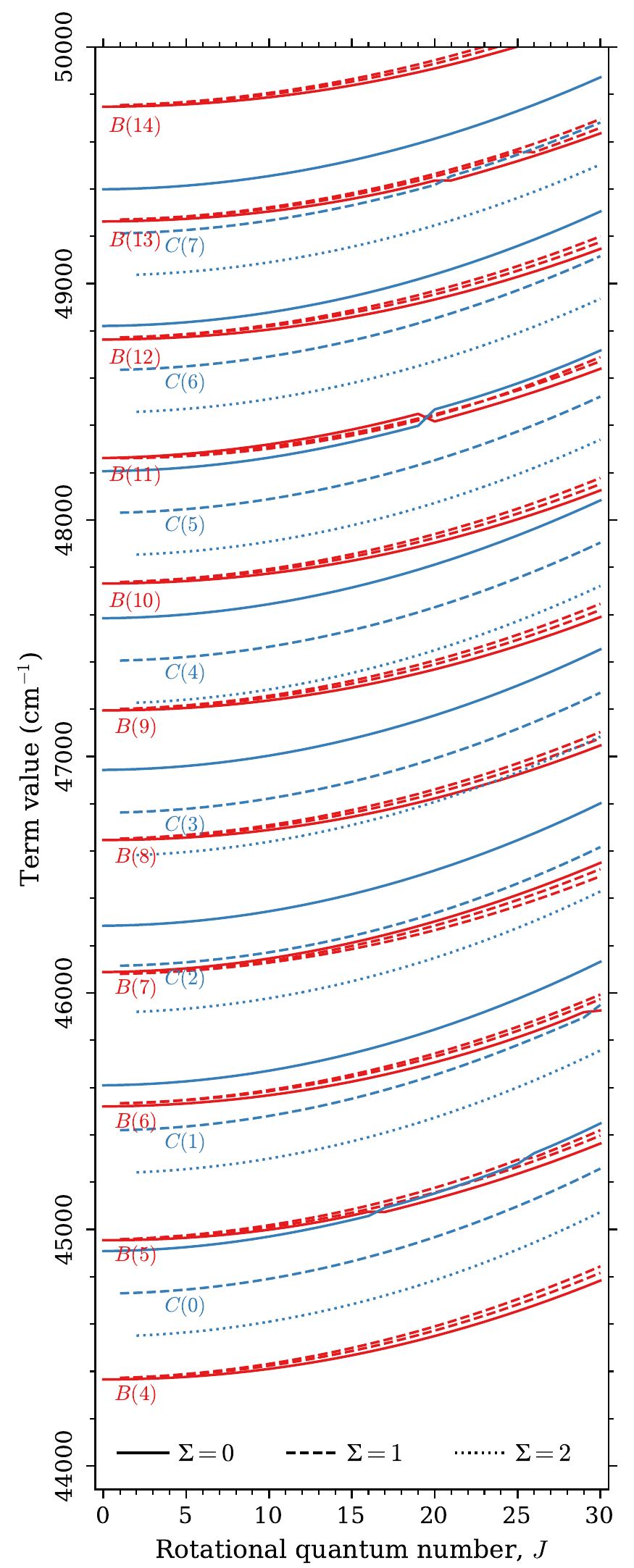}
  \caption{Observed or inferred \ce{{}^{32}S {}^{16}O} \Bstate and \Cstate levels.}
  \label{fig:all_levels}
\end{figure}

\subsection{Spectroscopic constants}
\label{sec:energy levels}

\begin{figure}
  \centering
  \includegraphics{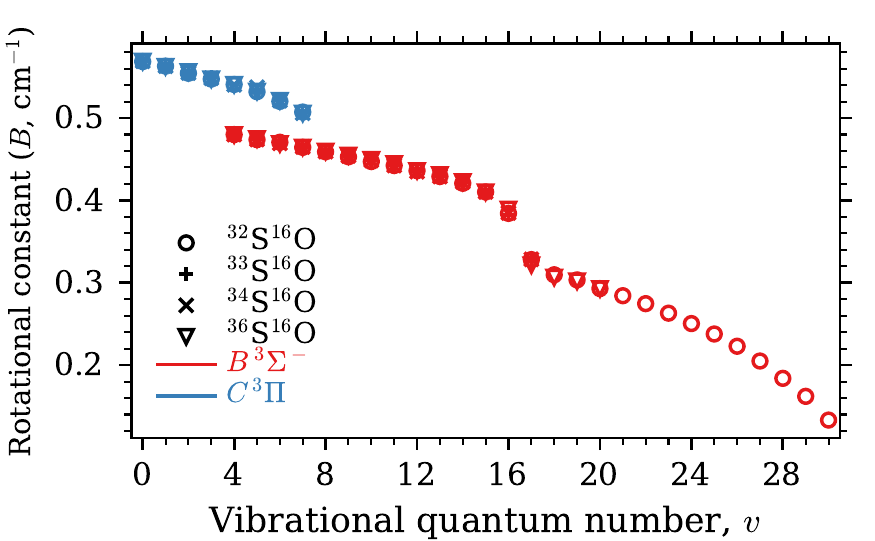}
  \caption{Fitted rotational constants of the observed $\Bstate(v)$ and $\Cstate(v)$ levels. The values for heavier isotopologues are reduced-mass-scaled upwards for comparison with \ce{{}^{32}S {}^{16}O}.}
  \label{fig:rotational_constants}
\end{figure}

\begin{figure}
  \centering
  \includegraphics{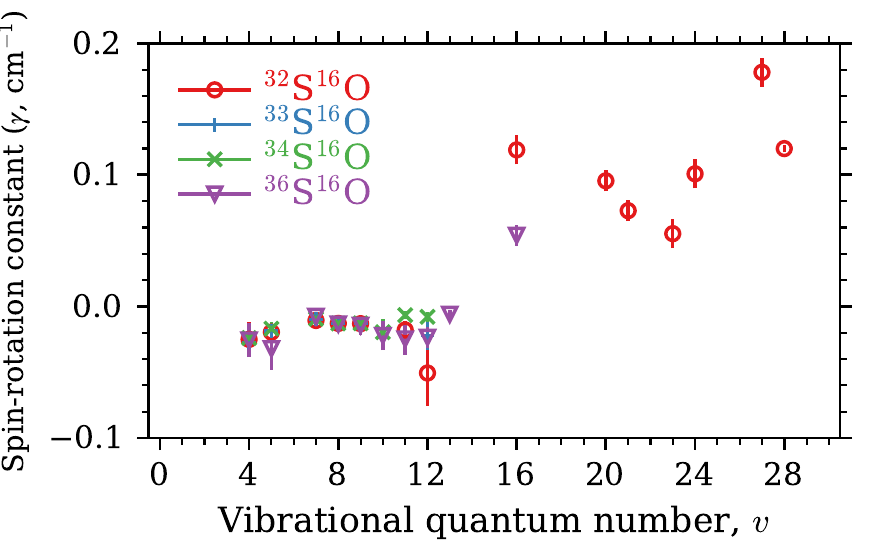}
  \caption{Fitted $\Bstate(v)$ spin-rotation constants.}
  \label{fig:gamma_constants}
\end{figure}

The fitted Hamiltonian parameters describing deperturbed energy levels are listed in Table~\ref{tab:level parameters}.
The fitted $B(v_B)\sim C(v_C)$ spin-orbit interaction parameters are plotted in Fig.~\ref{fig:interactions} and a level-energy map of fitted \ce{{}^{32}S {}^{16}O} levels up to $J=30$ and $B(v=14)$ is plotted in Fig.~\ref{fig:all_levels} showing various near-degeneracies and crossings of the \Bstate and \Cstate rotational series.

Parameters for some particularly broadened or perturbed bands could not be determined independently and were fixed to values in line with the spectrum as a whole and tabulated without uncertainties.
No spin-rotation interaction energies, $\ensuremath{\gamma}$, could be determined for any \Cstate levels.
Outlying values of $D$, $\ensuremath{\lambda}$, and $\ensuremath{\gamma}$ are mostly associated with absorption bands suffering from a combination of broadening, weakness, or contaminant overlap.

The rotational constants of \Bstate and \Cstate levels scaled by their reduced-mass are plotted in Fig.~\ref{fig:rotational_constants} and are in good agreement for the studied isotopologues.
A discontinuity between $B(16)$ and $B(17)$ is associated with the \Bstate outer limb irregularity depicted in Fig.~\ref{fig:introduction PECs}.
This is also coincident with a change in the measured spin-rotation parameter, $\ensuremath{\gamma}$, from approximately \wn{-0.02} for $B(v\leq 15)$ to about \wn{0.1} above, as shown in Fig.~\ref{fig:gamma_constants}.
The larger value may be identifiable with the higher-lying \ensuremath{{}^3\Sigma^-} state ``3'' computed by \citet{sarka2019b} and shown to exchange electronic character with \Bstate at large $v$, although the $\ensuremath{\gamma}$-constant was not calculated in that study.
The large scatter of fitted $\ensuremath{\gamma}$ values for $v\geq 16$ is indicative of significant model error in the band-by-band parameterisation of the congested spectrum approaching the \Bstate dissociation threshold, although the overall increase noted above is robustly measured.
A discontinuity of measured spin-spin interaction energies, $\ensuremath{\lambda}$, near $B(v=15)$ is not observed.

The fitted $C(v)$ spin-orbit constants, $A$, fall in the range $-182$ to \wn{-179.5}, and compare well with the $A=-181.4 \pm 0.1\wn{}$ values deduced by \citet{liu_ching-ping2006} from a well-resolved spectrum of \ce{{}^{32}S {}^{16}O} \CX{0}.
The approximate \wn{1} scatter of values is due to the broadening and weakness of some $C(v,\ensuremath{\Omega})$ levels, in which case they are correlated with both $T$ and $\ensuremath{\lambda}$ when fixing bandhead positions of the three $\ensuremath{\Omega}$-sublevels, with further correlation with spin-orbit interactions connecting nearby \Bstate levels.
Fixing all the $T$ and $\ensuremath{\lambda}$ parameters of \Cstate levels, along with their spin-orbit interaction with nearby \Bstate levels, to completely uncorrelated values requires a clear observation of transitions to all three $\ensuremath{\Omega}$-levels and their rotational structure.

Significant spin-orbit interactions are found to couple $B(v=5-8,10-13)$ levels in all isotopologues with the nearest $C(v)$ level, with rotational term series crossings and near-crossings for \ce{{}^{32}S {}^{16}O} shown in Fig.~\ref{fig:all_levels}.
No assumptions were made when fitting spin-orbit interaction energies, and their similarity across isotopologues and smooth dependence with increasing vibrational quantum numbers is evident in Fig.~\ref{fig:interactions}. 
The interaction energies mixing $B(13)$ with $C(7)$ for the various isotopologues are more scattered than for the mixing of lower-energy levels, and larger than might be anticipated from a simple extrapolation of vibrational excitation.
This apparent inconsistency is likely attributable to the particular difficulty in fitting the broadened $B(13)$ and $C(7)$ states and the weakness of absorption due to \CX{7} absorption.
In a simple test, a trial re-fitting of the \ce{{}^{32}S {}^{16}O} experimental spectrum was made assuming a $\ensuremath{\xi}=\wn{-8}$ interaction energy, consistent with an extrapolation of lower-$v$ interactions, and resulted in only a marginal reduction of its quality of fit.

\citet{liu_ching-ping2006} list $T$ and $B$ parameters fitted to their \ce{{}^{32}S {}^{16}O} spectra of $B(v=0-16)$ that are in reasonable agreement with our results.
However, there are large differences with respect to $D$, $\ensuremath{\lambda}$, and $\ensuremath{\gamma}$ constants, with those of Liu et al. including large magnitude and sign inconsistencies.
This is no doubt due to the difficulty of fitting rotational structure to the low-sensitivity and uncertain-excitation \BX{v} absorption spectra of \citet{liu_ching-ping2006}, which also prevented a deperturbation with respect to interacting $C(v>0)$ levels, as was possible here.
In general, the present $B(v>3)$ constants should be preferred.
A superior sensitivity to $\CX{0}$ in the combined spectra of \citet{liu_ching-ping2006} means that their fitted $C(0)$ parameters including a mutual deperturbation with both $d(1)$ and $B(5)$ are the more complete.
Their analysis includes several rotationally-mediated interaction terms mixing $C(0)$ and $B(5)$ which were not found necessary in our analysis of various $B(v_B)\sim C(v_C)$ couplings.
We find all $B\sim C$ perturbations to be well described by a single spin-orbit interaction energy, and the inclusion of rotational mixing of the order found by Liu et al. would be quite noticeable in our spectral modelling of the well-resolved $B(8)$ and $C(3)$ levels.
The cause of this difference is not clear but may arise from a greater sensitivity to $C(0)$ and  higher-$J$ $B(5)$ levels in the combined spectra available to Liu et al., leading to a good definition of level crossings near the  $J=21$ and 28, with fitted rotational interactions being more sensitive to higher-$J$ crossings.

\subsection{Potential-energy curves}
\label{sec:potential-energy curves}

\begin{figure}
  \centering
  \includegraphics{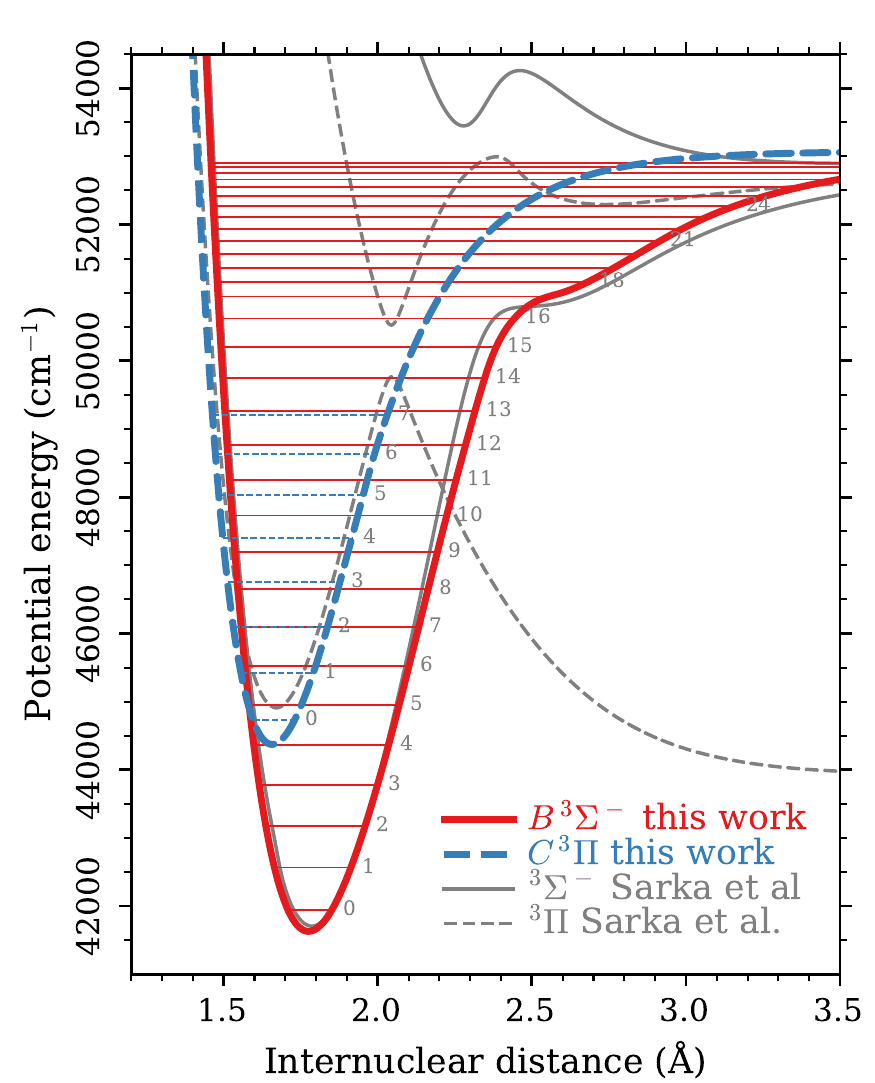}
  \caption{Empirical diabatic potential-energy curves for $(\ensuremath{\Omega}=0,J=0)$ \Bstate and \Cstate states relative to the ground-state equilibrium energy, and \ce{{}^{32}S {}^{16}O} vibrational level energies. Also shown are comparable adiabatic \abinitio curves computed by \citet{sarka2019b}.}
  \label{fig:PECs}
\end{figure}

The rotational level energies fitted to experimental spectra were used to deduce experimental diabatic \Bstate and \Cstate potential-energy curves and a spin-orbit interaction energy mixing the two electronic states.
The \Bstate experimental data were reinforced with \ce{{}^{32}S {}^{16}O}  $v=0$ to 3 levels computed from the molecular constants given by \citet{liu_ching-ping2006} to better constrain the \Bstate potential-energy minimum, but neglecting their perturbation by levels of \Astate. 
The solution of uncoupled vibrational energy levels and wavefunctions and the subsequent matrix diagonalisation to compute a perturbed spectrum are described in Sec.~\ref{sec:electronic state model}.

\begin{table}
  \begin{minipage}{\linewidth}
    \renewcommand{\thefootnote}{\textit{\alph{footnote}}}
    \caption{Morse potential-energy wells and interaction parameters.\protect\footnotemark[1]}
    \footnotetext[1]{The potential well formula: \begin{equation*} T(R) = T_\text{e} + c_2\left[1-\exp(-\ensuremath{\beta}(R-R_e))\right]^2, \end{equation*} applies to the given $v$-range. Higher-energy potential-energy curves are given numerically in the supplementary material. Uncertainties estimated by the least-squares optimisation of potential-energy curves and state interactions are given in parentheses in terms of the least-significant digit.}
    \label{tab:global_model_parameters}
    \begin{center}
      \begin{tabular}{c}
        \toprule
        \multicolumn{1}{c}{\Large$\Bstate$}\\[1ex]
        (Valid for $v\leq 14$)\\
        $T_e = \wn{41633.0(5)}$,
        $c_2  = \wn{20204.3(6)}$,\\
        $R_e = \np[\AA]{1.77526(5)}$, $\ensuremath{\beta}  = \np[\AA^{-1}]{+1.7634(3)}$, \\
        $\ensuremath{\lambda} = \wn{3.05(8)}$.\\
        \\
        \multicolumn{1}{c}{\Large$\Cstate$}\\[1ex]
        (Valid for $v\leq 4$)\\
        $T_e = \wn{44372.8(4)}$,
        $c_2  =\wn{14215.4(2)}$,\\
        $R_e = \np[\AA]{1.65708(6)}$, $\ensuremath{\beta}  = \np[\AA^{-1}]{2.372(2)}$, \\
        $A = \wn{-180.50(8)}$, \\
        $\ensuremath{\lambda} = \wn{1.35(7)}$  \\
        \\
        $\ensuremath{\xi}_{BC} = \wn{-56(4)}$ \\
        \bottomrule
      \end{tabular}
    \end{center}
  \end{minipage}
\end{table}

Fitted parameters governing this model are listed in Table~\ref{tab:global_model_parameters} and the standard deviation of residual differences between globally-computed rotational energy levels and those fitted band-by-band to the experimental spectra is \wn{1.3}.
The resulting potential-energy curves corresponding to $T(R)$ in Table~\ref{tab:electronic state matrix elements} are plotted in Fig.~\ref{fig:PECs}.
For each curve, the well and inner limb is fitted to one Morse function \citep{morse1929} and the outer limb described by another.
These regions are joined by a cubic-spline-interpolated region consisting of 9 spline knots between 2.1 and \np{2.7}{\AA} spanning the \Bstate potential inflection, and 2 knots at 1.88 and \np{1.95}{\AA} for \Cstate.
The \Cstate dissociation energy is unconstrained by our measurements and kept fixed at a value corresponding to the \ensuremath{\ce{S}({}^1{\rm D}_2)}+\ensuremath{\ce{O}({}^3{\rm P}_1)} excited atomic limit and relative to the ground-state dissociation energy deduced by \citet{clerbaux1994}.
The \Bstate dissociation limit was adjusted to best fit the experiment and its fitted value, \wn{53020}, lies between the \ensuremath{\ce{S}({}^1{\rm D}_2)}+\ensuremath{\ce{O}({}^3{\rm P}_1)} and \ensuremath{\ce{S}({}^1{\rm D}_2)}+\ensuremath{\ce{O}({}^3{\rm P}_2)} limits, but is poorly constrained given the uncertain appearance of $B(v>28)$ levels in our spectra.
All potential-energy curves are provided in numerical form in the supplementary material \citep{online_appendix}.

Spin-spin interaction energies for \Bstate and \Cstate, $\ensuremath{\lambda}$ in Table~\ref{tab:global_model_parameters}, were optimised in order to model all $\ensuremath{\Omega}$-levels and are in agreement with their values deduced band-by-band, as is the \Cstate spin-orbit splitting.
The spin-orbit interaction energy, $\ensuremath{\xi}_{BC}$, was fitted to an $R$-independent value of $-56 \pm 4$\wn{}, where the uncertainty is estimated by testing alternative values.
Specific $B(v_B)\sim C(v_C)$ interaction energies are computed from $\ensuremath{\xi}_{BC}$ and shown in Fig.~\ref{fig:interactions} to be in good agreement with the band-by-band interaction energies fitted to the experimental spectra.
Absolute magnitudes of the $R$-dependent $B\sim C$ spin-orbit interaction energies are calculated \abinitio by \citet{archer2000} and \citet{yu2011} and are 60 and \wn{35}, respectively, near 1.6\,\AA{} where the \Bstate and \Cstate potential-energy curves cross.
The former value is in good agreement with the present results.

\subsection{Transition moments}
\label{sec:transition moments}

\begin{figure*}
  \centering
  \includegraphics[width=\columnwidth]{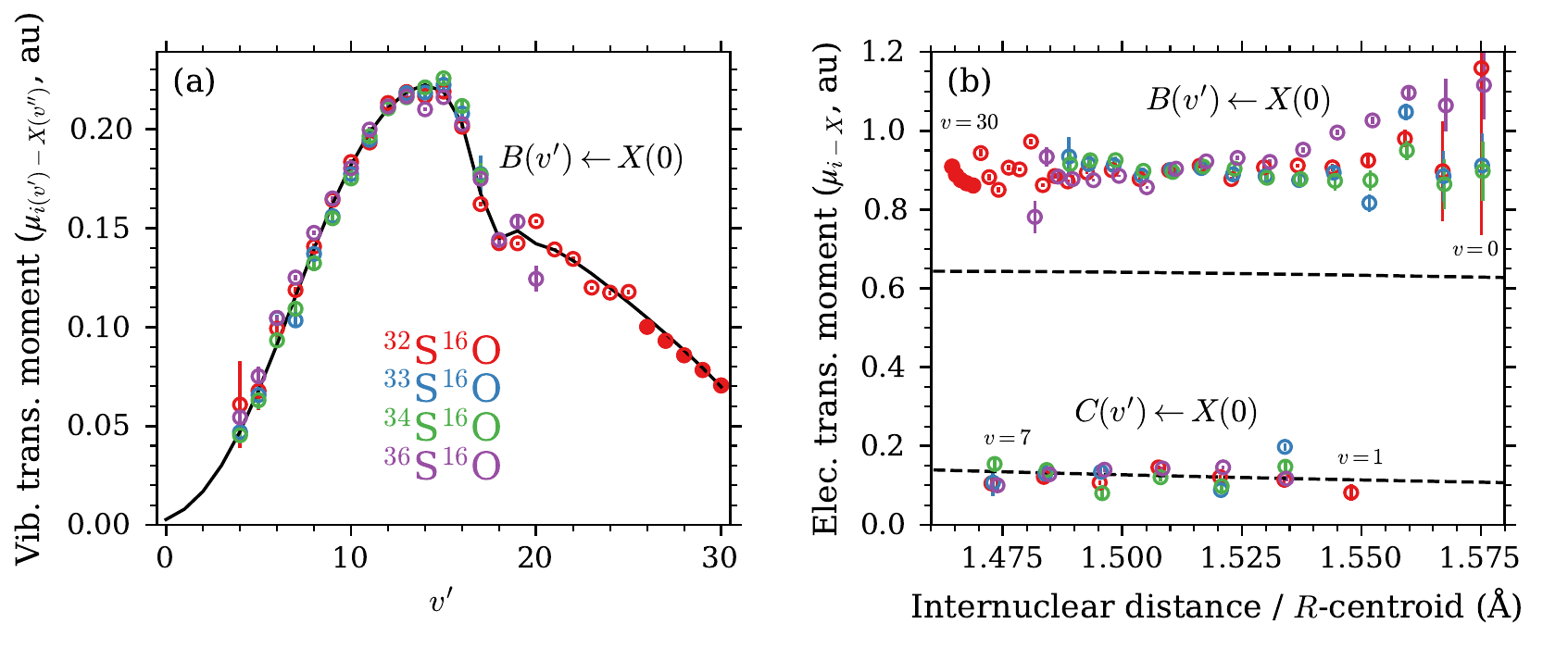}
  \caption{(a) Vibrational $B(v)-X(0)$ transition moments fitted band-by-band (error bars) and computed for \ce{{}^{32}S {}^{16}O} from the electronic-state model (curve). (b) Electronic transition moment deduced from vibronic values (error bars) and computed \abinitio by \protect\citet{sarka2019b} (curves).}
  \label{fig:transition moments}
\end{figure*}

Vibronic transition moments, $\ensuremath{\mu}_{i(v)-X(0)}$, deduced from the experimentally-observed \BX{v} and \CX{v} bands are listed in Table~\ref{tab:electronic state matrix elements} and plotted in Fig.~\ref{fig:transition moments} for \BX{v}.
The absolute scaling of these dipole moments was determined by reference to the observed \AX{v} spectrum with an estimated 10\% uncertainty, as discussed in Sec.~\ref{sec:A-X}.
The product of signs of the $\ensuremath{\mu}_{B(v_B)-X(0)}$, $\ensuremath{\mu}_{C(v_C)-X(0)}$, and $\ensuremath{\xi}_{B(v_B)C_(v_C)}$ parameters significantly affects the simulated spectrum of $B(v_B)-X(0)$ and $C(v_C)-X(0)$, and we find this product to be negative for all interacting bands observed here.
The uniquely-signed member of each triple is not determinable from our experimental data but may be identified in an \abinitio calculation computing matrix elements with self-consistent phase.
\citet{feng2019b} and \citet{sarka2019b} simultaneously compute $\ensuremath{\mu}_{BX}$ and $\ensuremath{\mu}_{CX}$ but find opposite and common signs, respectively.  
Here, we assume both transition moments to be positive, defining the spin-orbit interactions as negative.

Vibrational wavefunctions of $X(v=0)$, $B(v)$, and $C(v)$ levels were computed from a ground-state potential-energy curve generated by the Rydberg-Klein-Rees method \citep{rees1947} from data in \citet{lattanzi2015} along with the excited-state potential-energy curves described in Sec.~\ref{sec:potential-energy curves}. 
Franck-Condon factors and $R$-centroids \citep{lefebvre-brion_field2004} for the $B(v')-X(v'')$ and $C(v')-X(v'')$ transitions, computed using these wavefunctions are used to factor out the vibrational-dependence of experimental $\ensuremath{\mu}_{i(v)-X(0)}$ values, and the resulting $R$-centroid dependent $\ensuremath{\mu}_{B-X}$ and $\ensuremath{\mu}_{C-X}$ electronic transition moments are plotted in Fig.~\ref{fig:transition moments}.
We judge these to be $R$-independent within experimental uncertainty, with mean values $\ensuremath{\mu}_{B-X}=0.9\pm 0.1$ and $\ensuremath{\mu}_{C-X}=0.12 \pm 0.02 \,\text{au}$, where the estimated uncertainties are a combination of the 10\% calibration uncertainty and approximately 0.02\,au scatter of the experimental data in Fig.~\ref{fig:transition moments}.
There is a systematic overestimate of high-$v$ \ce{{}^{36}S {}^{16}O} \BX{v} transition moments that could not be eliminated satisfactorily, even with a biased refit of the \ce{{}^{36}S {}^{16}O} spectrum and its overlapping contaminants. This distortion is thus unexplained, but may arise from the ``jitter''  effect occasionally affecting SOLEIL FTS spectra and leading to an incorrect zero-intensity level \citep{de_oliveira2016}.

The present $B-X$ electronic transition moment for SO lies between the values determined experimentally for the analogous transitions in the isovalent molecules O$_2$ \citep{lewis_etal2001} and S$_2$ \citep{stark2018,lewis2018}, $\sim 0.87$ and $0.97 \pm 0.05$, respectively.
On the other hand, $B - X$ electronic transition moments for SO calculated \abinitio \citep{sarka2019b,feng2019b,da_silva2020} are significantly smaller than our experimental value in the region of internuclear distance probed by absorption from $X(v=0)$.
The cause of this difference is not clear.
Representative $R$-dependent \abinitio $B-X$ and $C-X$ electronic transition moments calculated by \citet{sarka2019b} are plotted in Fig.~\ref{fig:transition moments}, together with our experimental determinations.
While the large discrepancy in the $B-X$ case is evident, very good agreement occurs for the $C-X$ transition.

\begin{table}
  \caption{Radiative lifetime (ns) of $\protect\Bstate(v=0-3)$.}
  \small
  \label{tab:Bstate_lifetimes}
  \centering
    \begin{tabular}{cccc}
      \toprule
      $v$ & This work & \citet{elks1999} & \citet{yamasaki2005}\\
      \midrule
      0  &$37(4)$  & $33.6(6)$  & $29(2)$ \\
      1  &$37(4)$  & $32.3(6)$  & $30(4)$ \\
      2  &$38(4)$  & $36.4(5)$  & $27(4)$ \\
      3  &$37(4)$  & $52(2)~~$& $29(4)$     \\
      \bottomrule
    \end{tabular}
\end{table}

Our scaled electronic transition moments were further assessed by comparison with emission lifetimes measured previously for the unpredissociated $B(v=0-3)$  levels.
For this we computed vibrationally-averaged emission rates corresponding to $B(v'=0-3)\rightarrow X(v''=0-30)$ according to Eq.~(\ref{eq:band emission rate}) and computed the total $B(v')$ emission lifetime by summing over $v''$.
The results are listed in Table~\ref{tab:Bstate_lifetimes} with uncertainties following from the estimated 11\% uncertainty of our deduced $B-X$ transition moment.
Our predicted lifetimes are 10\% and 30\% longer than the time-domain measurements of \citet{elks1999} and Yamasaki et al. \citep{yamasaki2003b,yamasaki2005}, respectively (ignoring the outlying $B(3)$ lifetime of \citet{elks1999}).

\subsection{Predissociation broadening}
\label{sec:linewidths}
\begin{figure*}
  \centering
  \includegraphics[width=\columnwidth]{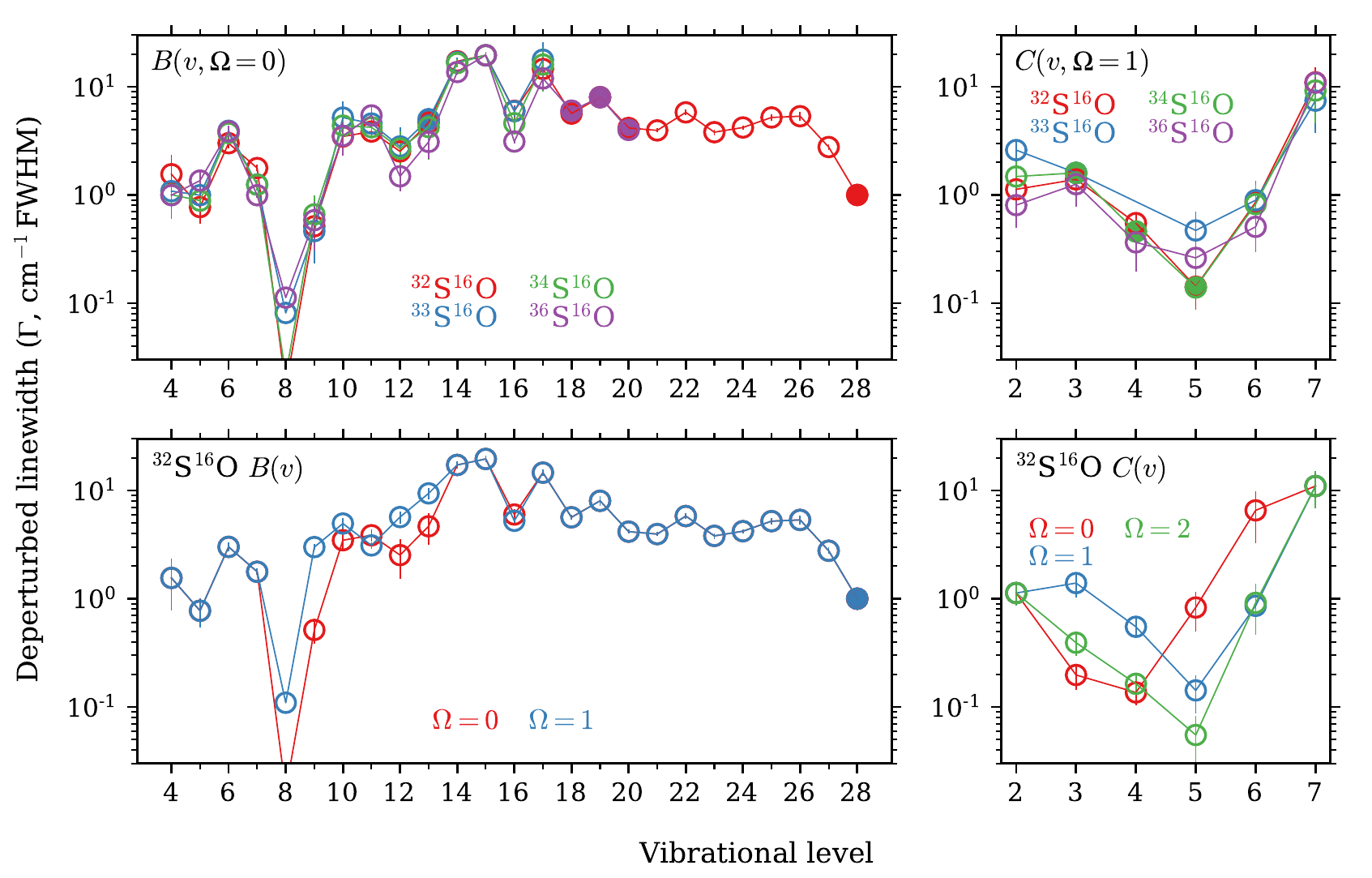}
  \caption{Deperturbed predissociation linewidths of the observed $B(v)$ and $C(v)$ states.
    The rotation-dependent widths of $B(8)$ are shown extrapolated to $J=0$.  \emph{Open symbols:} Fitted parameters. \emph{Closed symbols:} Fixed to assumed values.
    \emph{Upper figures:} Widths of $\ensuremath{\Omega}=0$ sublevels for all isotopologues. \emph{Lower figures:} Widths of all $\ensuremath{\Omega}$ levels for \ce{{}^{32}S {}^{16}O} only.}
  \label{fig:linewidths}
\end{figure*}

The deperturbed line broadening fitted to observed \BX{v} and \CX{v} bands is plotted in Fig.~\ref{fig:linewidths} and shows significant vibrational, isotopologue and $\ensuremath{\Omega}$ dependence.
The deperturbed widths for $\ensuremath{\Omega}=0$ and 1 $e$-parity \Bstate levels become rapidly mixed with increasing rotation so that the observed widths of high-$J$ levels converge.

The upper limits of \ce{{}^{32}S {}^{16}O} $B(v)$ widths estimated by \citet{liu_ching-ping2006} are in agreement with the values determined here, apart from their limits $\ensuremath{\Gamma}_{B(6)}<1.2$ and $\ensuremath{\Gamma}_{B(11)}<\np[cm^{-1}]{1}$ that fall below our measured widths of $(3.0\pm{}0.3)$\wnfwhm{} and $3.1-\wnfwhm{3.9}$, respectively.

The modelled \Bstate predissociation width pattern is likely caused by spin-orbit interaction between \Bstate and various crossing unbound states known from quantum-chemical calculations.
\citet{yu2011} compute all singlet, triplet, and quintet states dissociating to ground state \ensuremath{\ce{S}({}^3{\rm P})} and \ensuremath{\ce{O}({}^3{\rm P})} atoms which may provide such predissociation channels.
They find significant interaction energies mixing \Bstate with their $(1)\,{}^5\ensuremath{\Pi}$ and $(2)\,{}^5\ensuremath{\Pi}$ states, and the outer limb of an adiabatic $C\,{}^3\ensuremath{\Pi}$ state.
The bound $(1)\,{}^5\ensuremath{\Pi}$ state crosses \Bstate below the $\ensuremath{\ce{S}({}^3{\rm P})}+\ensuremath{\ce{O}({}^3{\rm P})}$ dissociation limit and provides a threshold dissociation channel beginning with $B(v=4)$.
This crossing likely also explains the predissociation of rotationally-excited $B(v=0-3)$  with their $J$-thresholds known experimentally \citep{martin1932,clerbaux1994}.
The $(2)\,{}^5\ensuremath{\Pi}$ state is predicted to cross \Bstate near $v=13$ and could provide an explanation for the rapidly increasing predissociation widths of higher levels.
An outer-limb crossing with the adiabatic $C\,{}^3\ensuremath{\Pi}$ state may further enhance \Bstate predissociation near $v=10$.
The rapid variation of width with $v$ between threshold and these crossings likely results from the varying overlap of bound- and unbound-state radial wavefunctions, as is typical of predissociation by outer-limb crossing repulsive states \citep{lefebvre-brion_field2004}. 
\citet{yu2011} predict additional crossings and spin-orbit interactions between \Bstate near $v=7$ and 11, and the repulsive outer limbs of two \ensuremath{{}^1\Pi} states.
This purely $\ensuremath{\Omega}=1$ interaction could explain the enhanced $\Omega=1$ dissociation widths we find for $B(9)$ and $B(10)$.
Finally, we note that the $J$- and $\Omega$-dependencies of the $B(v=8)$ linewidths for \ce{{}^{32}S {}^{16}O}, displayed in Fig.~\ref{fig:spectrum B08 widths}, are typically characteristic of predissociation by a ${}^3\Pi$ state \citep{julienne_krauss1975}, also observed in some levels of the $B$ state of \ce{O2} \citep{lewis1994}.
Thus, the \Cstate state is likely to be partially involved in the, possibly complex, $B(v=8)$ predissociation mechanism.

Our analysis of $C(v)$ levels reveals a broad width minimum around $v=4$ and up to factor-of-five differences between $\ensuremath{\Omega}$-substates of the same vibrational level, with either $\ensuremath{\Omega}=0$ or 1 being the most broadened.
The large widths of $C(v=6,\ensuremath{\Omega}=0)$ and $C(v=7)$ are consistent with being nearest to the \ensuremath{{}^3\Pi} avoided crossing shown in Fig.~\ref{fig:PECs}.

\subsection{Rotational line lists and cross sections}

\begin{figure*}
  \centering
  \includegraphics[width=\columnwidth]{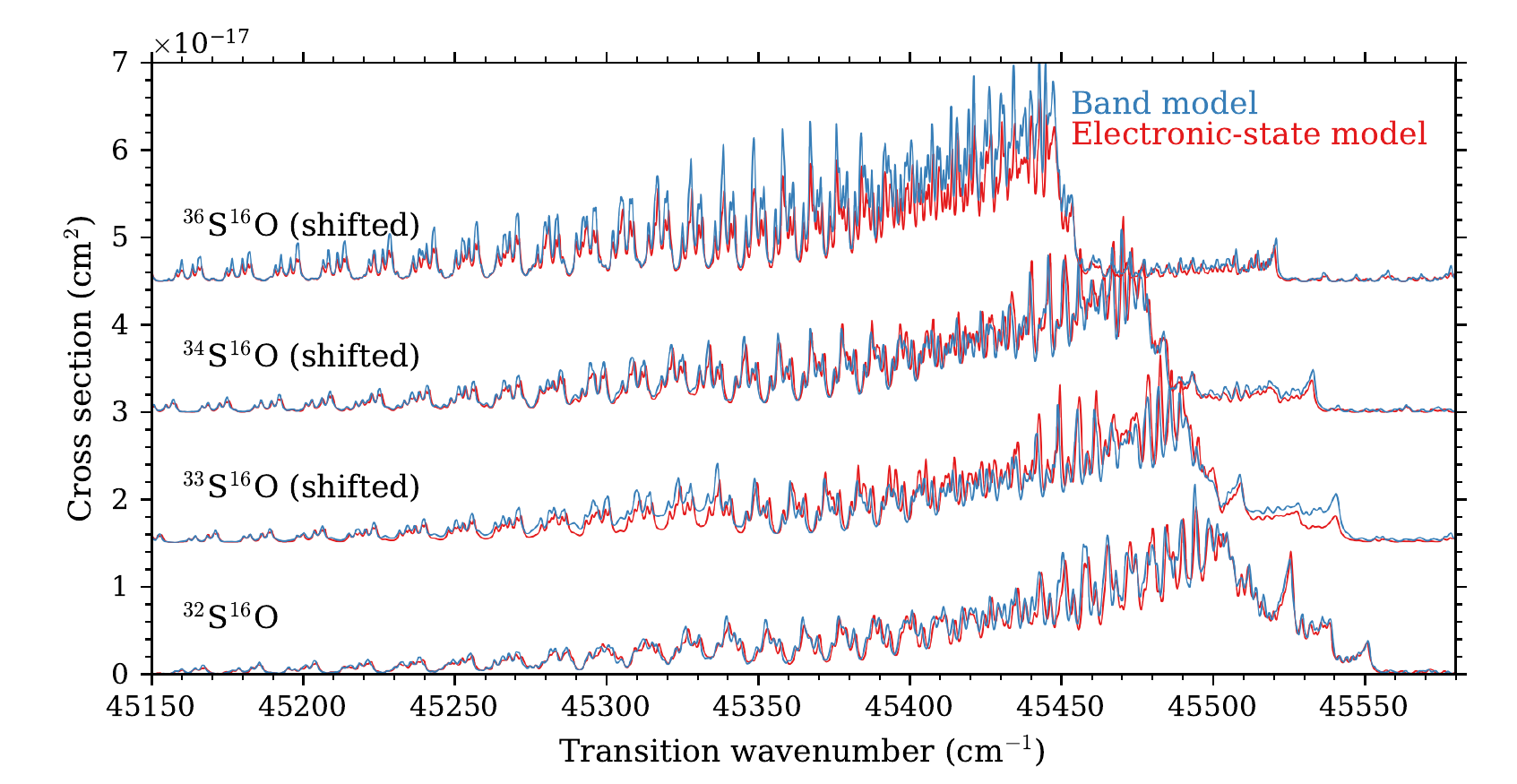}
  \caption{Simulations of absorption into the mixed $B(7)\sim C(2)$ levels computed from constants fitted band-by-band and a global electronic-state potential-energy-curve model.}
  \label{fig:spectrum_mod_exp_comparison_B07}
\end{figure*}

\begin{figure*}
  \centering
  \includegraphics[width=\columnwidth]{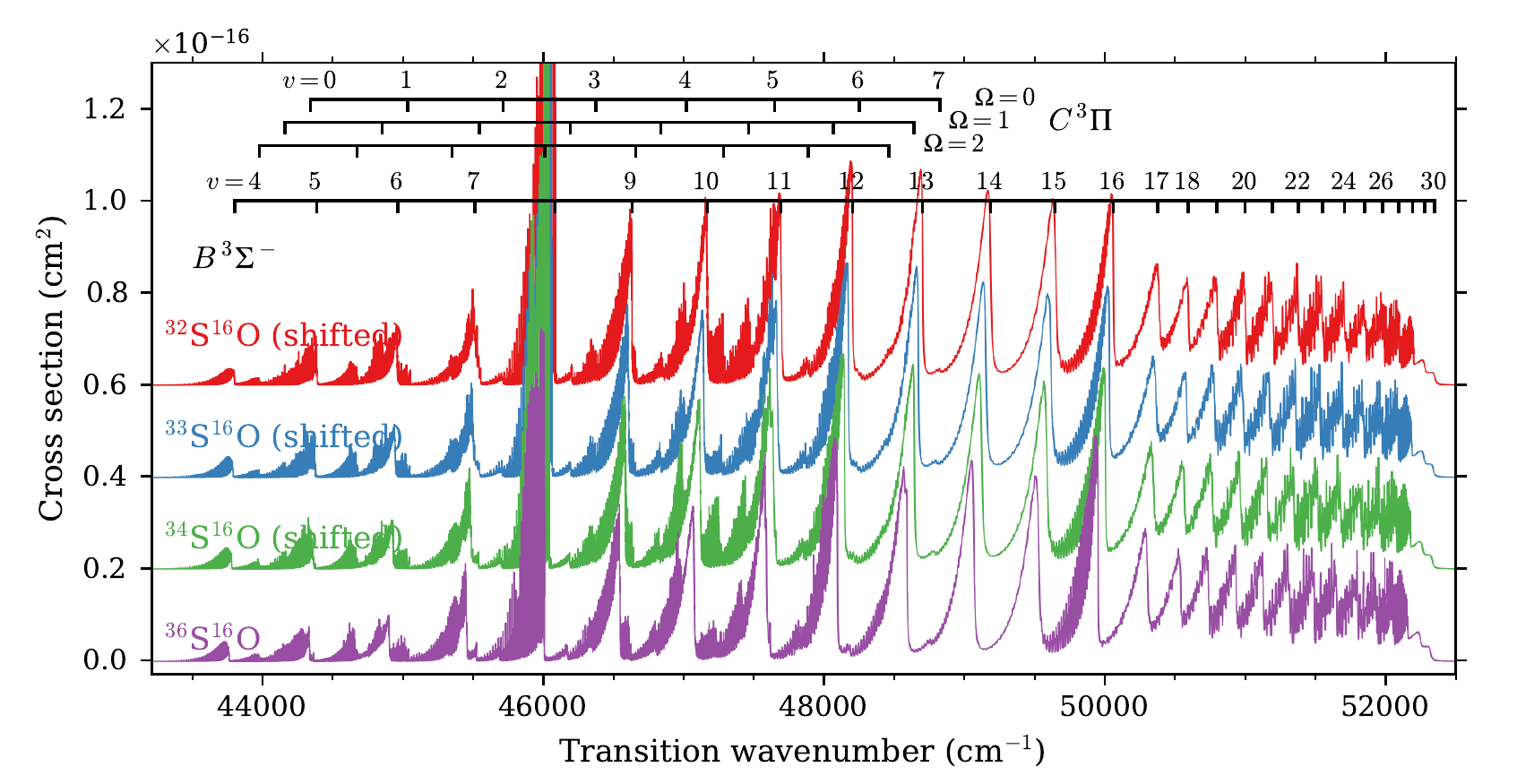}
  \caption{Synthetic photodissociation spectra computed at 360\,K.}
  \label{fig:spectrum overview}
\end{figure*}

The band-by-band modelling detailed in Sec.~\ref{sec:results} results in a rotational line list for all observed bands, while the global model described in Secs.~\ref{sec:potential-energy curves} and \ref{sec:transition moments} also permits the calculation of line frequencies and strengths for bands that are not experimentally observed in some isotopologues.
We combine band-by-band and global-model derived line lists in order to generate a complete line list that is as accurate as possible.

A global-model line list of rotational transitions terminating on $B(v=0-30)$ and $C(v=0-7)$ levels up to $J=50
$ was computed using the potential-energy curves, electronic state interactions, and transition moments presented in Secs.~\ref{sec:potential-energy curves} and \ref{sec:transition moments} for all measured isotopologues and, for good measure, \ce{{}^{18}O}-substituted species.
Band-by-band-fitted deperturbed linewidths are added to the global-model line list.
The unmeasured linewidths of S-substituted isotopologues for $B(v>16)$ levels, and all levels of O-substituted isotopologues are set to their \ce{{}^{32}S {}^{16}O} values.

A cross section for absorption into the strongly-coupled $B(7)\sim C(2)$ levels is computed from the band-by-band constants fitted to the experimental spectra, as well as from the global-model line list.
These are compared at 360\,K in Fig.~\ref{fig:spectrum_mod_exp_comparison_B07}, and the detailed rotational structure and mass-dependent spin-orbit mixing of levels is well reproduced by the global model.
The cross sections of other SO bands show similar agreement when making this comparison, with the principal differences being due to band-strength differences and small frequency shifts.
The globally-modelled strengths smooth over the background intensity and contaminant absorption uncertainties affecting band-by-band analysis, and are therefore more accurate.
For example, the band-by-band \ce{{}^{36}S {}^{16}O} cross section in Fig.~\ref{fig:spectrum_mod_exp_comparison_B07} is clearly an overestimate when compared with the isotopically self-consistent global model.
Conversely, the band-by-band modelled line frequencies of observed bands are more accurate than the global model.
We construct a final hybrid line list by substituting band-by-band measured line frequencies into the electronic-state model.
Photoabsorption cross sections computed from this hybrid list for all S-isotopes are shown in Fig.~\ref{fig:spectrum overview}.

Absorption into $B(v>3)$ and all $C(v)$ levels is completely dissociative given the short lifetimes implied by the large measured transition widths. 
Additionally, levels of \ce{{}^{32}S {}^{16}O} $B(v=3)$ with quantum-number $N\geq 10 $ are known to dissociate, with only lower-$N$ levels contributing to the emission spectra of \ce{{}^{32}S {}^{16}O} \citep{martin1932,clerbaux1994}.
Emission spectra of S-substituted isotopologues have not been measured and will likely have higher-$N$ predissociation thresholds, as found for \ce{{}^{12}S ^{18}O} by \citet{clerbaux1994}.
We neglect this difference given the rather small contribution of $B(3)$ to the overall photodissociation cross section and adopt the $N=10$ threshold for all species.
Then we compute a photodissociation cross section from the hybrid line list by including only dissociative upper levels.

Band-by-band and global model rotational line lists, and the recommended hybrid of the two, are permanently available in an online data archive \citep{online_appendix}.

\subsection{Comparison with other published cross sections}
\label{sec:comparison with previous cross sections}

\begin{figure}
  \centering
  \includegraphics{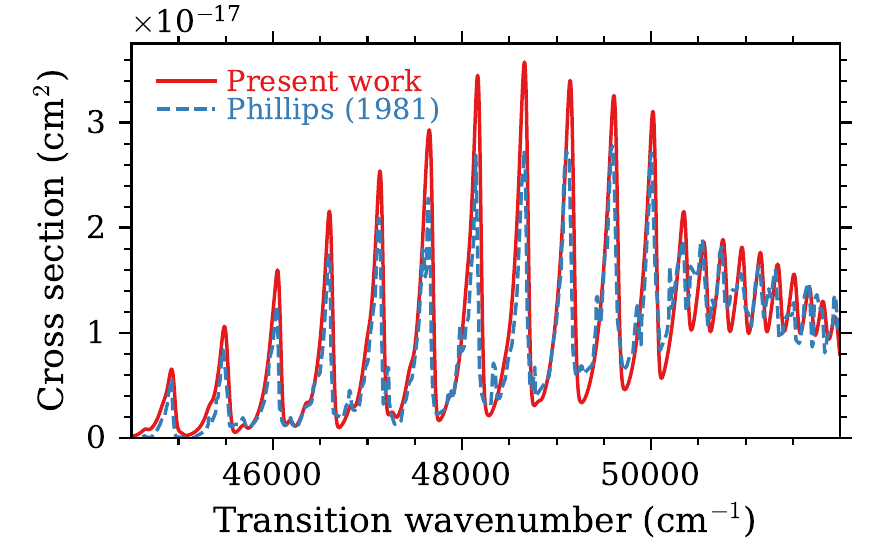}
  \caption{The present \ce{{}^{32}S {}^{16}O} cross section computed assuming a rotational temperature of 400\,K and degraded by convolution with a unit Gaussian of width \wnfwhm{50}, compared with the low-resolution photoabsorption cross section of \citet{phillips1981}.}
  \label{fig:compare cross section}
\end{figure}

\citet{phillips1981} measured the \BX{v} cross section for \ce{{}^{32}S {}^{16}O} at low spectral resolution and a digitised version is plotted in Fig.~\ref{fig:compare cross section}.
We compare this with a \ce{{}^{32}S {}^{16}O} cross section computed from our final hybrid line list assuming a rotational temperature of 400\,K and broadened by convolution with a Gaussian function of width \wnfwhm{50}, in order to approximately match the \citet{phillips1981} instrumentally-broadened band profiles.
The relative band strengths are similar and integrated cross sections over the region plotted agree within 15\%, with the new cross section being larger.

There are absorption features appearing in the cross section of \citet{phillips1981} which are not found in our analysis of SO spectra, for example at \np{48400} and \wn{52400}.
These are broadly aligned with \ce{SO2} absorption bandheads and likely result from an imperfect subtraction of contaminant \ce{SO2}.
\citet{danielache2014} make a similar comparison of their cross section computed \abinitio with \citet{phillips1981}, finding reasonable relative agreement but with some differences in relative \BX{v} band strengths and frequencies, but, more importantly, the \abinitio cross section was found to have a three-times smaller magnitude overall.

\section{Summary}

Detailed spectroscopic data on the $\Bstate(v=4-30)$ and $\Cstate(v=0-7)$ states of SO and  its S-substituted isotopologues are determined from high-resolution absorption spectra covering the \np{43000} to \wn{51000} (190 to 233\,nm) spectral region.
This is the first observation of \Cstate levels above $v=2$ or a \Bstate or \Cstate level in any S-substituted isotopologue.

Most of the observed bands are severely blended due to a high line density and predissociation broadening, and their profiles were fitted to minimally-specified effective Hamiltonians including spin-orbit interactions between neighbouring $B(v_B)$ and $C(v_C)$ levels.
The fitted linewidths are quite variable with respect to vibrational level and quantum number $\ensuremath{\Omega}$ and likely arise from further interactions with unbound electronic states.
An empirical model of \Bstate and \Cstate electronic states was constructed to ensure globally-reliable band strengths and to enable extrapolation to unmeasured vibrational levels and isotopologues.

The fitted cross section has a band-by-band relative uncertainty within 5\% and an additional 10\% absolute uncertainty based on published lifetime and radiative data, and their uncertainties, for the $\Astate(v=1)$ level \citep{lo1988,elks1999}. 
The estimated total uncertainty encompasses agreement with a previously measured \ce{{}^{32}S {}^{16}O} cross section \citep{phillips1981} and $\Bstate(v=0-2)$ lifetimes \citep{elks1999}, but somewhat disagrees with a further set of $\Bstate(v=0-3)$ lifetimes \citep{yamasaki2005}.
However, the new $\Bstate-\Xstate$ transition moment is in significant disagreement with previous \abinitio calculations \citep{sarka2019b,da_silva2020} that suggest a 50\% smaller cross section, while $\Cstate-\Xstate$ transition moments are found to agree well.

The final result is a comprehensive and spectroscopically-accurate line list of photodissociating far-ultraviolet rovibronic transitions for all \ce{{}^{x}S{}^{16}O} isotopologues.
These data, available in an online archive \citep{online_appendix}, are ideally suited for computing SO lifetimes against photodissociation in atmospheres and the interstellar medium, and the resulting likelihood of photolytic S-isotope fractionation.

\section*{Acknowledgements}
The authors are very pleased to contribute to this special issue in honour of Prof. Wim Ubachs.
We have benefited enormously from his pioneering work in high-resolution and time-resolved spectroscopy and his insightful and energised collaboration on numerous projects.
We thank Bérenger Gans of the Institut des Sciences Moléculaires d'Orsay for permitting the use of the radio-frequency discharge source.
AH was funded by the NASA Postdoctoral Program through the NASA Astrobiology Institute, by grant number 19-03314S of the Czech Science Foundation, and the ERDF/ESF ``Centre of Advanced Applied Sciences'' (grant number CZ.02.1. 01/0.0/0.0/16\_019/0000778).
JRL acknowledges support from the NASA Exobiology program (grant \#80NSSC19K0475 to Arizona State University).
 

\end{document}